\documentstyle[epsfig]{mn2e}
\begin{document}
\input BoxedEPS.tex
\newcommand{\aip}{{\small ${\cal AIPS}$}}
\newcommand{\gtsim}{\mbox{{\raisebox{-0.4ex}{$\stackrel{>}{{\scriptstyle\sim}}
$}}}}
\newcommand{\ltsim}{\mbox{{\raisebox{-0.4ex}{$\stackrel{<}{{\scriptstyle\sim}}
$}}}}
\newcommand{\s}{$\stackrel{\rm s}{.}$}
\newcommand{\h}{$^{\rm h}$}
\newcommand{\m}{$^{\rm m}$}
\newcommand{\pp}{$\stackrel{\prime\prime}{.}$}
\newcommand{\de}{$^{\circ}$}
\newcommand{\p}{$^{\prime}$}
\newcommand{\arc}{$^{\prime\prime}$}
\newcommand{\marc}{^{\prime\prime}}
\newcommand{\rs}{{\em $r_s$}}
\newcommand{\DPM}{{\em DPM}}
\newcommand{\alf}{{\displaystyle\biggl({\nu_{\rm h} \over \nu_{\rm l}}\biggr)^{\alpha}} }

\newcommand{\figstart}[1]
    { \begin{figure}[htb]
      \begin{picture}(0,#1) }
\newcommand{\figend}[4]
    { \end{picture}
      \special{#1}
      \caption[#2]{#3}
      \label{#4}
      \end{figure} }
\newcommand{\fig}[5]
    { \figstart{#1}
      \figend{#2}{#3}{#4}{#5} }
\newcommand{\bHS}{\beta_{\mbox{\scriptsize HS}}}
\newcommand{\bBF}{\beta_{\mbox{\scriptsize BF}}}
\newcommand{\nT}{\nu_{\mbox{\scriptsize T}}}
\newcommand{\et}{E_{\mbox{\scriptsize T}}}
\newcommand{\nTn}{\nu_{\mbox{\scriptsize Tn}}}
\newcommand{\nTf}{\nu_{\mbox{\scriptsize Tf}}}
\newcommand{\tn}{\tau_{x\mbox{\scriptsize n}}}
\newcommand{\tf}{\tau_{x\mbox{\scriptsize f}}}
\newcommand{\xn}{x_{\mbox{\scriptsize n}}}
\newcommand{\xf}{x_{\mbox{\scriptsize f}}}
\newcommand{\yn}{y_{\mbox{\scriptsize n}}}
\newcommand{\yf}{y_{\mbox{\scriptsize f}}}
\newcommand{\lln}{l_{\mbox{\scriptsize n}}}
\newcommand{\llf}{l_{\mbox{\scriptsize f}}}
\newcommand{\Dn}{f(\Delta_{\mbox{\scriptsize n}})}
\newcommand{\Df}{f(\Delta_{\mbox{\scriptsize f}})}
\newcommand{\B}{\mbox{$B$}}
\newcommand{\Bo}{\mbox{$B$}_{0}}

\SetEPSFDirectory{/scratch/sbgs/figures/hst/}
\SetRokickiEPSFSpecial
\HideDisplacementBoxes

\title[AGN dust tori: the X-ray-infrared connection]{AGN dust tori: the X-ray-infrared connection}
\author[Rowan-Robinson M.]{Michael Rowan-Robinson$^1$, Ivan Valtchanov$^{1,2}$, Kirpal Nandra$^1$\\
$^1$ Astrophysics Group, Blackett Laboratory, Imperial College of Science 
Technology and Medicine, Prince Consort Road,London SW7 2AZ\\
$^2$  Herschel Science Centre, ESAC, ESA\\ 
}
\maketitle
\begin{abstract}
We have combined the well-studied CLASXS Chandra survey
in Lockman with the Spitzer SWIRE survey data to study the X-ray-infrared connection for AGN.  The sample consists
of 401 X-ray-sources, of which 306 are detected by Spitzer, and a further 257 AGN candidates detected through their dust
torus, but not by Chandra.  We have used spectroscopic redshifts and classifications from the literature, where available, 
and photometric redshifts for the remainder.  For X-ray 
sources the X-ray hardness ratio has been modelled 
in terms of a power-law ($\Gamma$ = 1.9) with absorption N(H).  The optical and infrared data have been modelled 
in terms of our well-established optical galaxy and QSO templates, and infrared templates based on radiative
transfer models.  This type of analysis  gives better 
insight into the infrared SEDs, and a better separation of the contribution of starbursts and AGN dust tori, than a simple 
comparison of 24 $\mu$m to optical or X-ray fluxes.  We also believe this gives more insight than using a library
of fixed UV-infrared templates. 

Our estimate of the N(H) distribution is consistent with other studies, but we do find a higher proportion of low absorption 
objects at z $<$ 0.5 than at z $>$ 0.5.  
While we find only one X-ray AGN with N(H) $> 10^{24} cm^{-2}$, we argue that 10 objects with torus luminosity apparently
exceeding the bolometric X-ray to 3$\mu$m luminosity are strong candidates for being heavily absorbed in X-rays.
We also estimate that at least half of the infrared-detected AGN dust tori which are undetected in X-rays are likely to be 
Compton thick.  Our estimate of the total number of Compton-thick objects in the 0.4 sq deg area is $\ge$ 130, 
corresponding to $\ge 20\%$ of the combined SWIRE-CLASXS sample (and with an upper limit of 39 $\%$).

We find no evidence for AGN with no dust tori, and none with a covering factor $<1\%$ but there are clear examples
of AGN with covering factors of only a few percent and these, though rare, do not fit easily with a unified picture
for AGN.  
The range of dust covering factors is 1-100 $\%$, with a mean of 40$\%$, ie a Type 2 fraction of 40$\%$.  
Measured by the ratio of dust torus luminosity to X-ray or (for Type 1 objects) optical luminosity, the covering factor
appears to decrease towards intermediate AGN luminosity, in contradiction to estimates based on ratios of narrow-line and
broad-line spectra, but may increase again at low AGN luminosity.   

We find 7-10 candidate X-ray starbursts in the SWIRE-CLASXS sample, with X-ray luminosities ranging up to 
$L_{Xh} = 10^{44}$ erg s$^{-1}$.
This is a considerable extension of the luminosity range of X-ray starbursts previously reported, but is consistent 
with the an extrapolation of the X-ray-infrared relation for starbursts into the realm of hyperluminous infrared galaxies.


\end{abstract}


\section{Introduction}

The idea that AGN are surrounded by dust which absorbs their visible and ultraviolet light and 
reemits it at mid infrared wavelengths was first put forward by Rowan-Robinson (1976).  He also
suggested that extinction by this dust was responsible for the distinction between Type 1 and 2
AGN, an idea which has developed into the unified model for AGN (Antonucci 1993, Krolik 1999).  IRAS showed
that a mid-infrared excess is a common feature of AGN (Miley et al 1984) and this was
interpreted in terms of a dust torus around the AGN nucleus (Rowan-Robinson and Crawford 1989,
Pier and Krolik 1992, Granato and Danese 1994, Efstathiou and Rowan-Robinson 1995).
Rowan-Robinson (1995) proposed that the 'torus' might be an axisymmetric distribution of
discrete, optically thick, clouds and gave a 1-dimensional treatment of the radiative transfer
for such an ensemble of clouds.  Recently 3-dimensional radiative transfer codes have been
applied to such cloud ensembles by Hoenig (2006) and Nenkova et al (2008a,b).

In this paper we explore the connection between X-ray emission from AGN and the mid-infrared emission
from the dust torus.  We take advantage of the fact that the wide-area CLASXS Chandra survey 
(Yang et al 2004, Steffen et al 2004) lies
within the SWIRE-Lockman area survey by Spitzer.  This allows us to construct complete samples of AGN
selected either in X-rays or at mid-infrared wavelengths, and to explore both the common sources and
those undetected in either waveband.  The X-ray and optical properties of the CLASXS survey have been
analyzed by Barger et al (2005).  The X-ray mid-infrared connection has been explored in  the SWIRE/Chandra
survey of a 0.6 sq deg northern Lockman field,  at similar X-ray sensitivity to CLASX, by Polletta et al (2006), and 
in a shallower survey of 2.6 sq deg in the XMM MDS by Tajer et al (2007).  Mid-infrared and X-ray spectral energy 
distributions (SEDs) of AGN in  the XMM MDS have been analyzed by Franceschini et al (2005) and Polletta et al (2007).  
In particular Polletta et al (2006)
have discovered several Compton-thick AGN with prominent dust tori but very weak optical emission. 
Compton-thick quasars have been studied recently by Daddi et al (2007), Fiore et al (2008, 2009) and Alexander et al (2008)
(see also Polletta et al 2006 for references to earlier work).
 
Our study is well-suited to address the issues of the prevalence of Compton thick AGN and whether 
there exists a major population of
AGN which are absorbed even at 10 keV but might still contribute a significant fraction of the hard X-ray
and infrared backgrounds.  This is of interest because about half of the X-ray background is unresolved at 6 keV 
(Worsley et al 2005).  Models have been proposed to explain the observed background which involve strong
dependence of the obscured AGN fraction on X-ray luminosity and redshift (eg La Franca et al 2005, Gilli et al 2007) or where
the obscured fraction is independent of luminosity and redshift (eg Treister and Urry 2005). Both types of model 
assume a high fraction of obscured AGN.
 We also explore whether there are AGN with very weak or absent dust tori
and the role of X-ray starbursts in the sample.

\section{The CLASXS Chandra survey}
The CLASXS survey is described by Yang et al (2004).  It covers an area of 0.4 sq deg in the Lockman 
hole to a Chandra on-axis X-ray sensitivity of  5x$10^{-16}$ erg s$^{-1}$ cm$^{-2}$ at 0.4-2 keV and 
3x$10^{-15}$ erg s$^{-1}$ cm$^{-2}$ at 2-8 keV .    
Optical photometry at BVRIz', spectroscopic redshifts and a spectroscopic classification are
described by Steffen et al (2004).  For sources in common with SWIRE we also have UgriZ photometry.
We have used the combined optical photometry to derive photometric redshifts, 
using the methodology of Rowan-Robinson et al (2008), for sources without spectroscopic redshifts.
Because we can be reasonably certain that most of these sources are AGN, we have dropped the
prior used in Rowan-Robinson et al (2008) that QSOs should be classified as optically starlike, and we have
also allowed extinction values, $A_V$ up to 3.0 for QSOs.

In the area in common with complete SWIRE coverage the CLASXS survey contains 424 hard X-ray sources 
of which 23 are stars identified spectroscopically.  

\subsection{Hydrogen column-density, N(H)}
We have followed Polletta et al (2006) in using the ratio of hard to soft X-ray count-rates to estimate N(H).
We used PIMMS3 to derive a look-up table for the ratio of ACIS-I (2-8 keV) hard to soft (0.4-2.0) band counts as a a function
of N(H) and redshift, assuming an underlying power-law continuum with $\Gamma$ = 1.9 (Nandra and Pounds 1994).   Tozzi et al (2006)
give a mean value for $\Gamma$ of 1.75, while Tueller et al (2008) give 1.98.  Where a
source is undetected in the soft X-ray band (6 sources), N(H) has been calculated assuming a soft count-rate
 of 0.02 and is a lower limit.  However we should bear in mind that Winter et al (2008) find that 60$\%$ of a
sample of hard X-ray sources have complex spectra, generally due to an additional soft X-ray excess, 
for which a simple power-law+absorption model will underestimate the true absorption.  On the other
hand Tozzi et al (2006), studying a sample similar to ours, find a smaller proportion (22$/$82 bright X-ray sources)
showing either a scattered soft component or reflection-dominated spectrum.

Figure 1 shows N(H) versus X-ray luminosity.  The X-ray luminosity, $L_{Xh}$, is the 4 keV rest-frame
monochromatic luminosity, corrected for the assumed N(H).  We have indicated the standard subdivisions of
N(H) into 'unabsorbed' (N(H) $< 10^{22} cm^{-2}$), 'obscured' ($10^{22} cm^{-2} < N(H) < 10^{24} cm^{-2}$),
or 'Compton thick' (N(H) $> 10^{24} cm^{-2}$), and the $10^{42}$ erg s$^{-1}$ luminosity limit above which 
X-ray sources are almost certainly AGN.  Figure 2L shows N(H) versus redshift.  
There is a strong selection effect against detecting high values of N(H) at low redshift (see Fig 14 of
Tozzi et al 2006).  High values of N(H) ($>10^{23.5} cm^{-2}$) can only be detected for z$>$2, since
at lower redshift the hard X-ray band would suffer serious absorption.  The N(H)-z distribution seen in
Fig 2L is similar to that found by Tozzi et al (2006). 
Fig 2R shows the same plot for X-ray sources with no SWIRE counterparts.  
The latter tend to be just a higher redshift subsample of the X-ray population.

Type 1 QSOs tend to have lower values of N(H), as expected in unified models, in which Type 1 QSOs are 
assumed to be viewed face-on.

\begin{figure*}
\epsfig{file=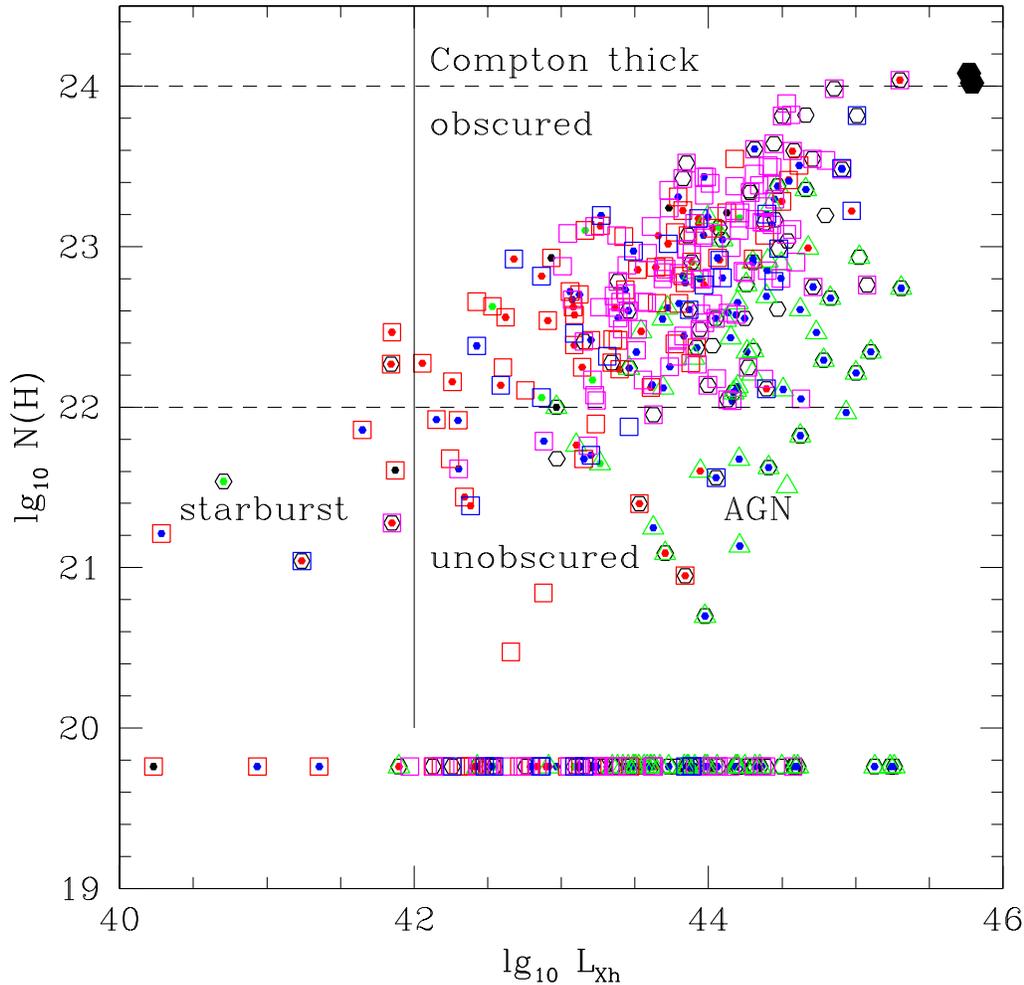,angle=0,width=14cm}
\caption{Hydrogen column-density, N(H), versus hard X-ray luminosity.
The colour and symbol coding in this and subsequent figures are as follows:
optical types (open symbols): black circle: optical QSO, green triangle: broad line, red square:
galaxy, blue square: Seyfert 2, magenta square: no spectral classification.
infrared template types (filled circles): black: cirrus, red: M82 starburst, green: Arp220 starburst,
blue: AGN dust torus.  The large filled 
circles are the two Compton-thick QSOs studied by Polletta et al (2006).
}
\end{figure*}

There are 88 sources for which the hardness ratio implies no absorption and these
have been assigned a galactic absorption of $10^{19.76} cm^{-2}$ (cf Yang et al 2004).

Figure 3 shows histograms of the N(H) distribution for redshift bins between z = 0 and 3.  While the shifting of the
high N(H) cutoff to higher values of N(H) as redshift increases is a selection effect as higher
absorption drops sources out of the hard X-ray sample at lower z, the dearth of the low N(H) values at higher
redshift seems to be a real effect.  Because at higher redshift there is poorer resolution of low values of N(H), it is 
better to simply consider the total fraction of relatively unabsorbed sources.  The percentage of low absorption 
(N(H) $< 10^{22} cm^{-2}$) X-ray sources
is 58, 37, 42, 32, 33, 39$\%$ for z = 0-0.5, 0.5-1, 1-1.5, 1.5-2, 2-2.5, 2.5-3, respectively, so the increase at z $<$ 0.5
seems real but there is no change in the fraction between z = 0.5 and 3.
Tueller et al (2008) find for their low redshift Swift BAT 14-195 keV sample a flat distribution of $log_{10}$ N(H) from 20.5 to
23.5.  Like Tozzi et al (2006), we seem to see fewer low N(H) systems in the CLASXS Chandra sample than Tueller et al.
\begin{figure*}
\epsfig{file=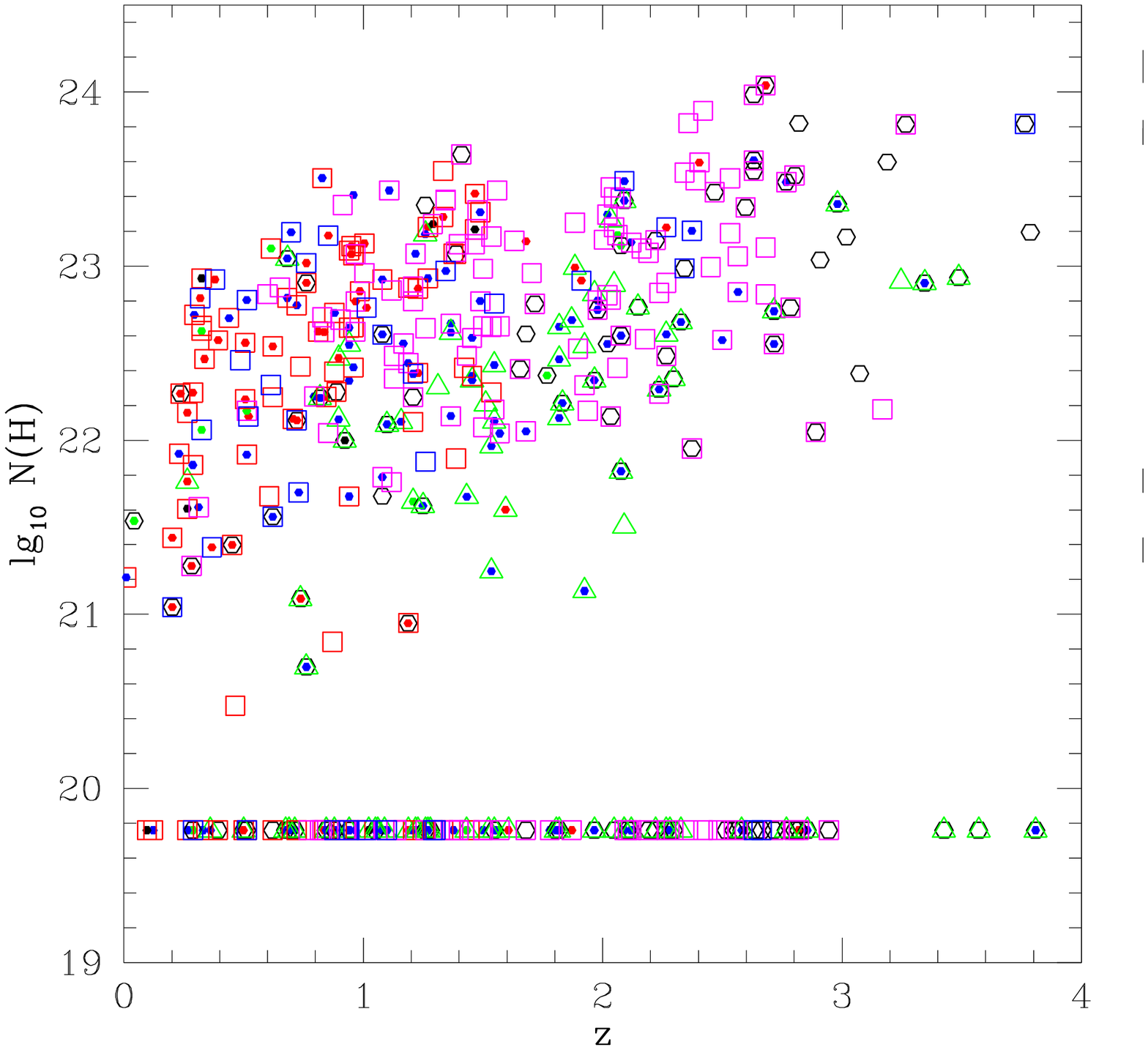,angle=0,width=7cm}
\epsfig{file=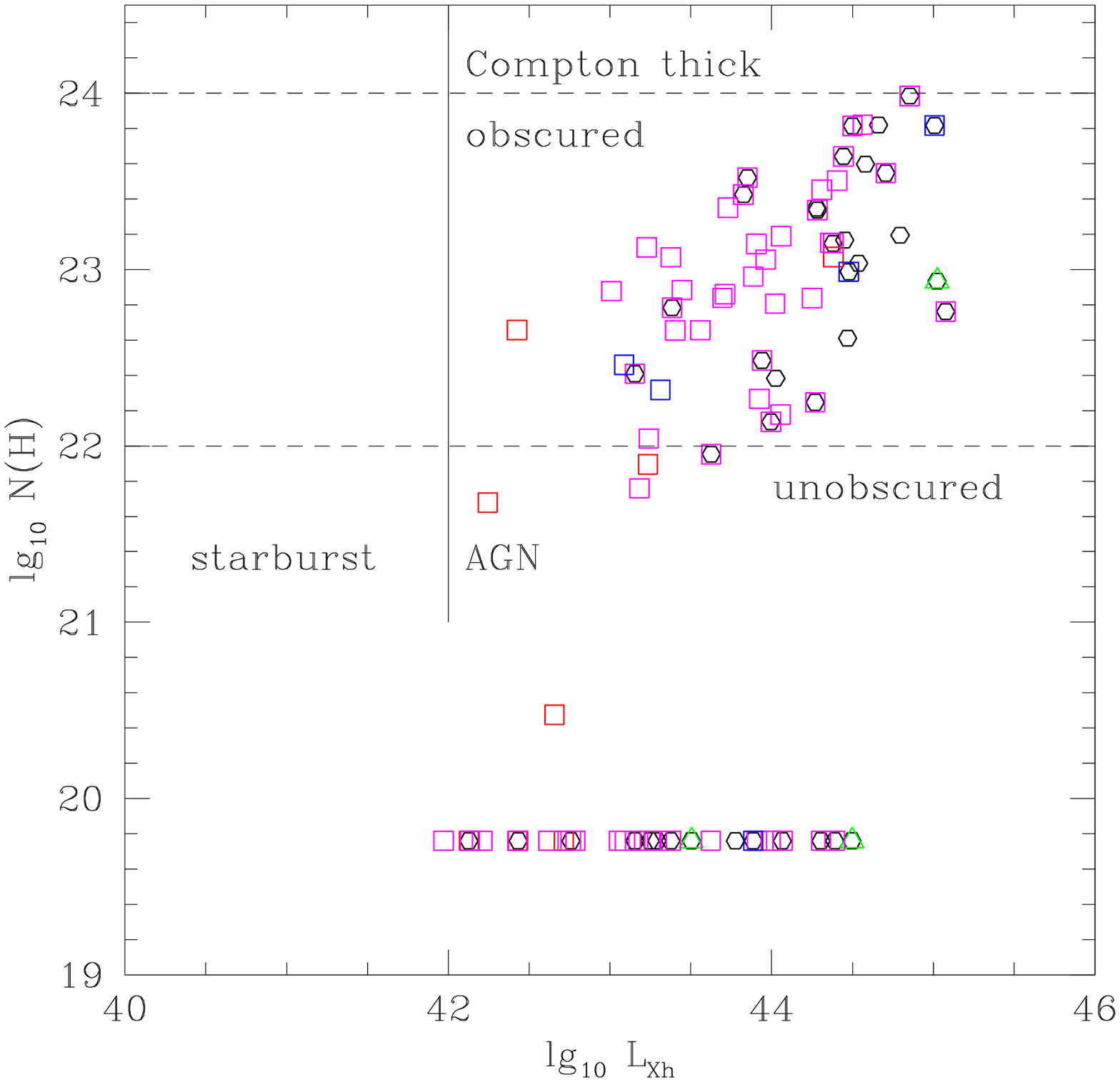,angle=0,width=7cm}
\caption{LH: Hydrogen column-density, N(H), versus redshift.  
RH: Hydrogen column-density, N(H), versus hard X-ray luminosity, 
for X-ray sources with no SWIRE counterparts.
}
\end{figure*}

\begin{figure*}
\epsfig{file=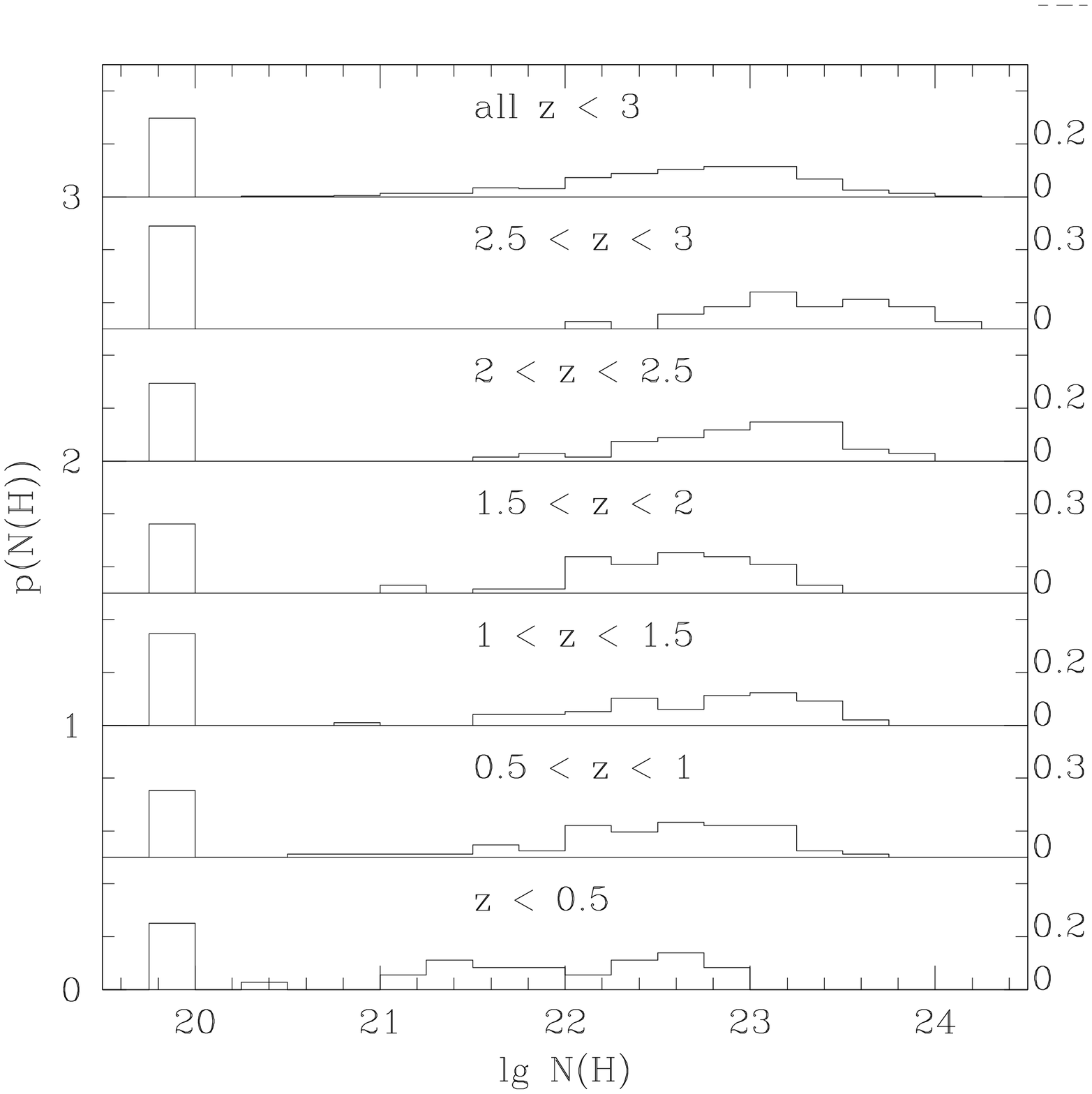,angle=0,width=14cm}
\caption{Histogram of the N(H) distribution for different redshift bins.  Objects for which the inferred N(H) $< 10^{20} cm^{-2}$
have all been assigned a value $10^{19.76}$}
\end{figure*}

\subsection{X-ray bolometric correction}
Figure 4L shows $L_{opt}$ versus $L_{Xh}$ for all the CLASXS sources.  Fig 4R shows the corresponding
plot just for objects with Type 1 QSO optical templates. For Type 1 QSOs we can make a reasonably accurate 
estimate of the bolometric (X-ray to near ir) luminosity, since the SED peaks at UV wavelengths.  We assume 
a bolometric correction to the 0.1-3 $\mu$m luminosity, $L_{opt}$, of 2.0 (Rowan-Robinson et al 2008).
Where we have only X-ray luminosity as an indicator of the AGN luminosity, the issue is more problematic.   
In Fig 4R we see a wide range of ratios of optical to X-ray luminosity for  Type 1 AGN, which translates to
a wide range of bolometric corrections in the X-ray bands.  Vasudevan and Fabian (2007) find 
bolometric corrections at 2-10 keV in the range 4-100.   Dependence of the bolometric correction on 
luminosity have been investigated by Marconi et al (2004), Steffen et al (2006), Shemmer et al (2008),
and on the Eddington fraction, $L/L_{Edd}$, by Vasudevan and Fabian (2007).  We find for the present sample 
(for optical QSOs with $log_{10} L_{Xh} > 43.0$) a mean value of 27 for the bolometric correction in the hard 
X-ray (4 keV) band and have used this value to estimate the bolometric (X-ray to near ir) luminosity, 
$L_{Xh,c}$, where necessary, but the large uncertainty in such estimates needs to be kept in mind.
We have not attempted to correct for a dependence of bolometric correction on luminosity.

\begin{figure*}
\epsfig{file=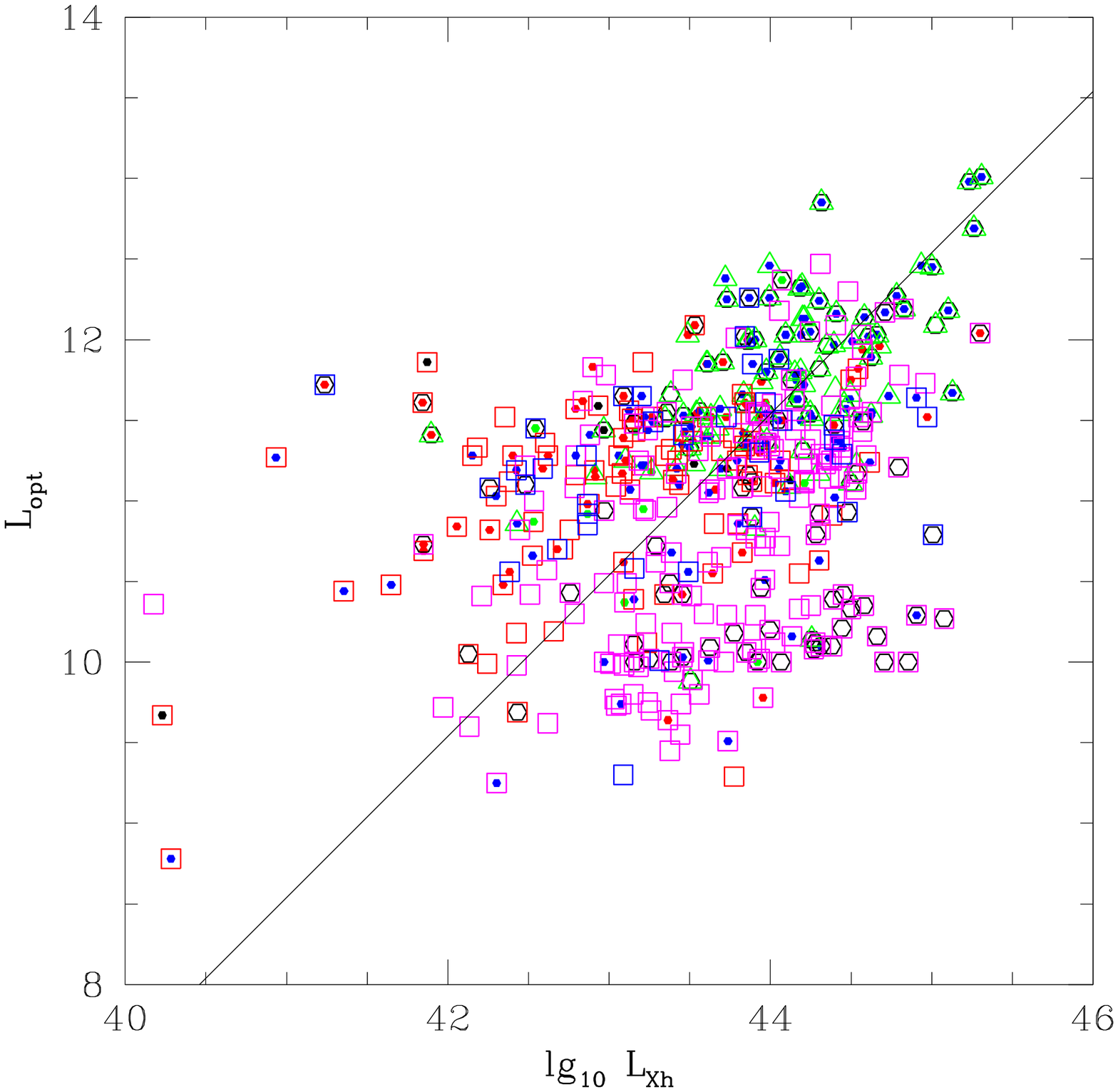,angle=0,width=7cm}
\epsfig{file=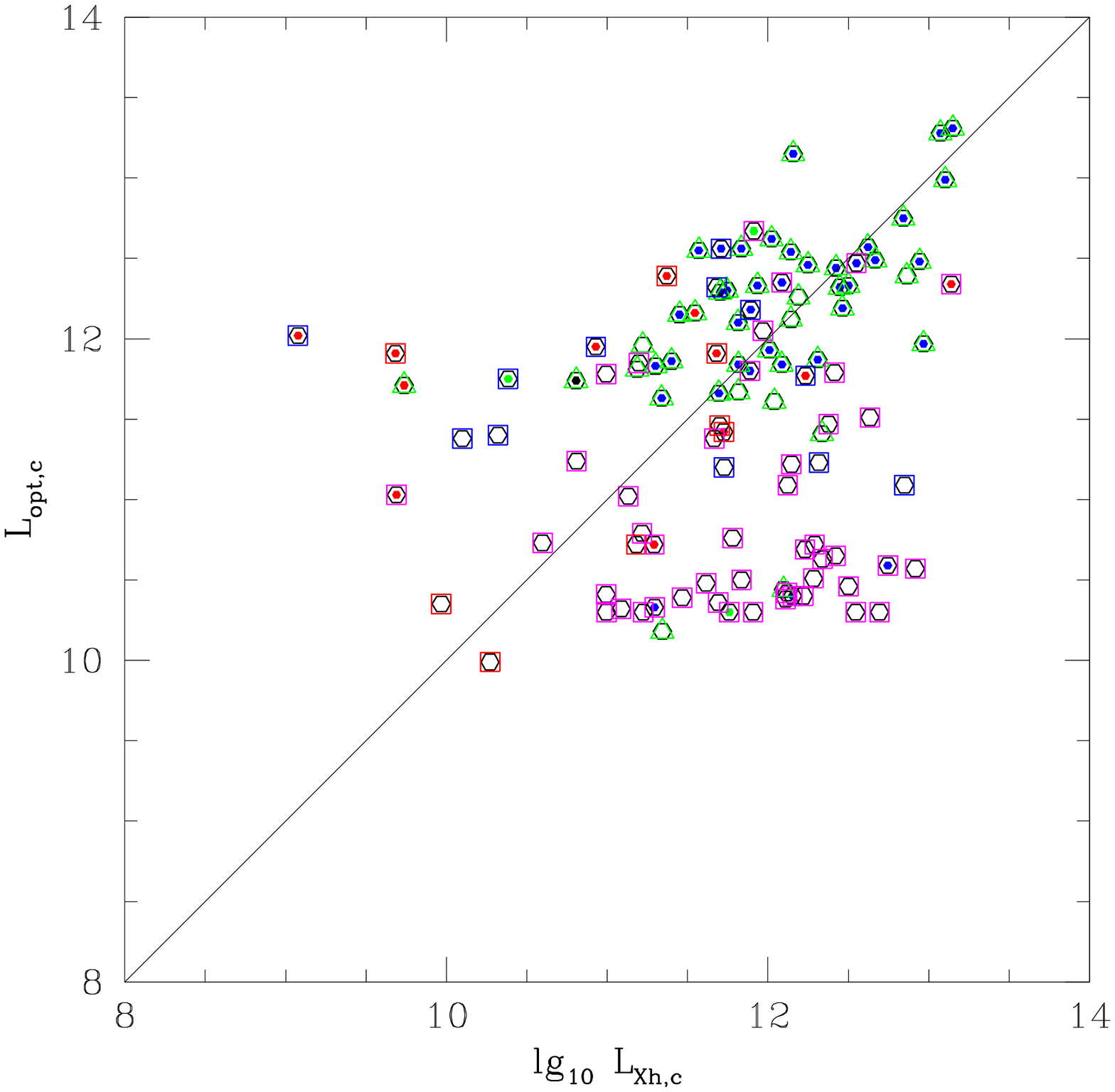,angle=0,width=7cm}
\caption{Optical luminosity, $L_{opt}$, versus X-ray luminosity, $L_X$.  LH plot: all sources, 
RH plot: optical QSOs only.  The solid line corresponds to equal bolometric luminosities assuming an
optical-UV bolometric correction of 2.0 and a hard X-ray correction of 27.
Crosses denote Compton-thick sources ($N(H) > 10^{24} cm^{-2}$).
}
\end{figure*}

We have investigated the outliers in Fig 4R.  In Fig 5L we replot Fig 4R as the ratio $L_{opt,c}/L_{Xh,c}$
versus the ratio of hard to soft X-ray counts.   QSOs with high values of $L_{opt}/L_{Xh}$ seem to be
cases where the optical continuum may be contaminated by light from the parent galaxy, since they are flagged
as optically extended.  The QSO fits involve quite high values of $A_V$ ($>$1), and this suggests possible 
aliasing between highly reddened QSOs and galaxy SEDs in the optical template fitting.   QSOs with high 
values of $L_{Xh}/L_{opt}$ 
are mostly high redshift QSOs, with modest reddening, but high X-ray absorption.  They seem to be Type 2 objects 
in which the optical light is scattered from the accretion disk into the line of sight.  In Fig 5R we exclude objects 
flagged as optically extended or having N(H)$>10^{23} cm^{-2}$.  There is now a much tighter correlation of
optical and X-ray luminosities, with the corrected ratio $L_{opt,c}/L_{Xh,c}$ lying between 0.25 and 10, still 
implying quite a wide range of hard X-ray bolometric corrections. 

Table 1(a) summarizes properties of the CLASXS X-ray sample in four bins of X-ray luminosity.

\begin{figure*}
\epsfig{file=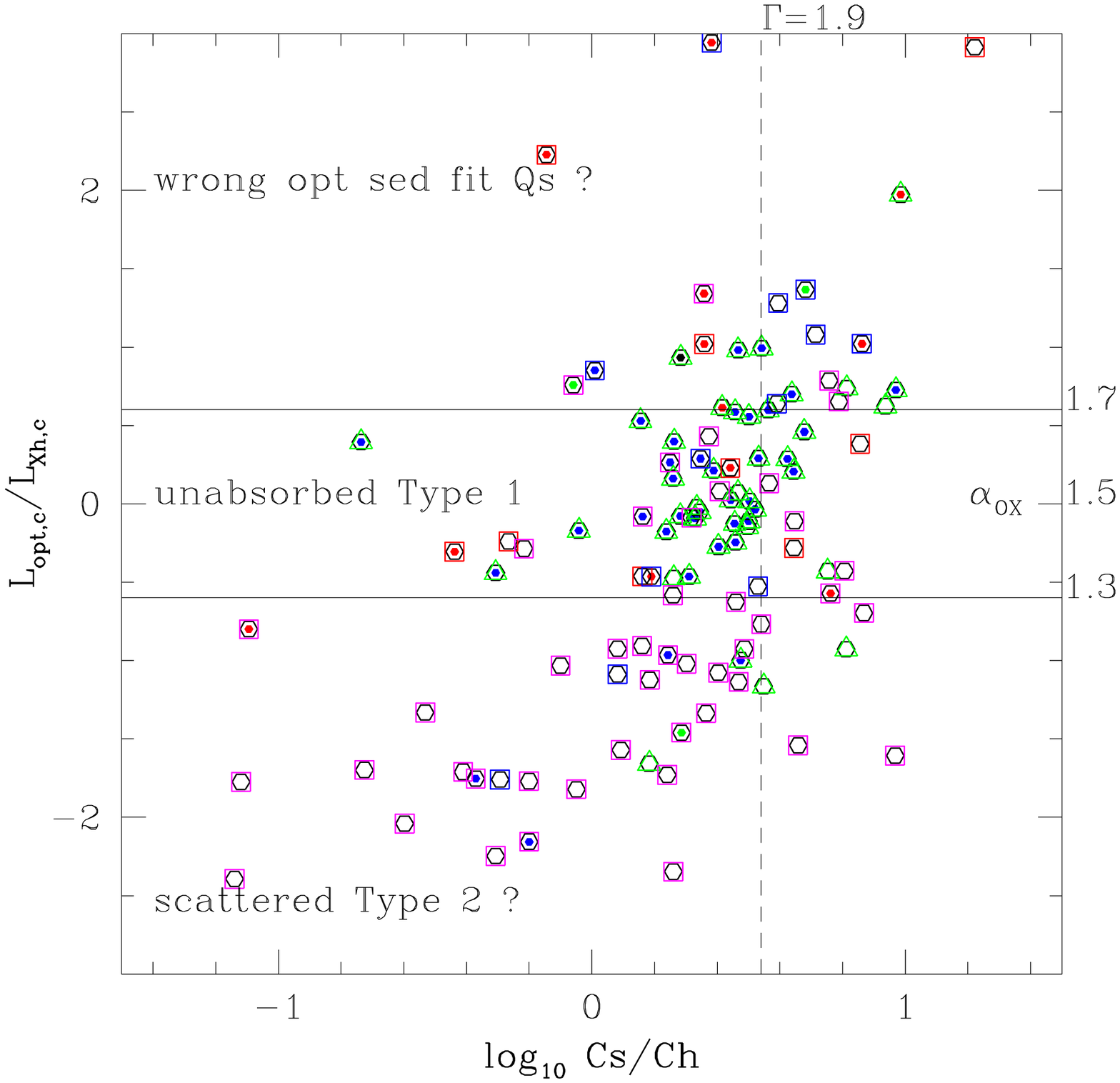,angle=0,width=7cm}
\epsfig{file=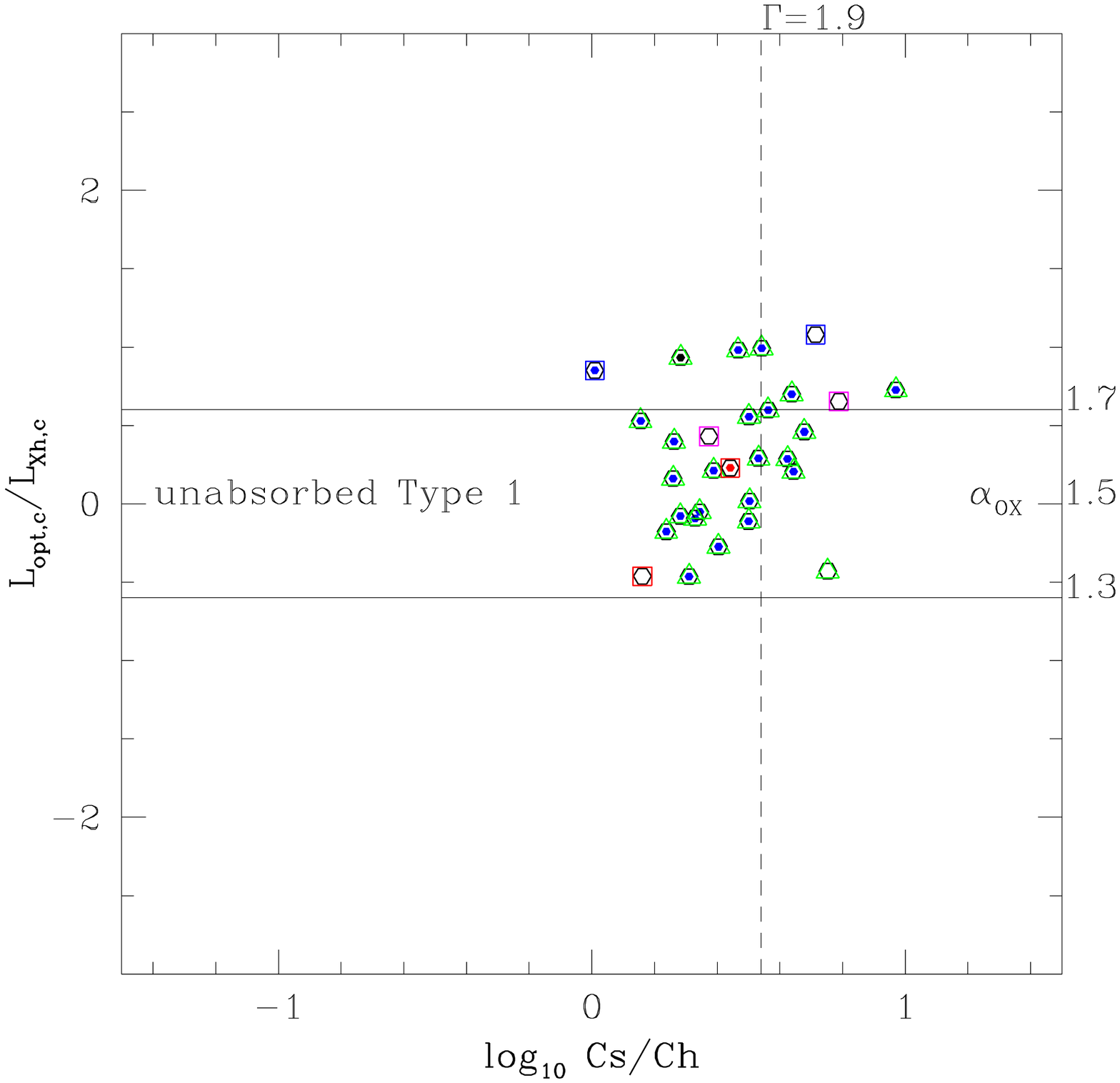,angle=0,width=7cm}
\caption{Ratio of optical luminosity, $L_{opt}$, to X-ray luminosity, $L_X$, versus ratio of hard to soft
X-ray counts.  LH plot: all sources assigned QSO templates, 
RH plot: Same, excluding QSOs flagged as optically extended or with N(H)$>10^{23} cm^{-2}$.  The solid lines 
correspond to $\alpha_{OX}$ = 1.3, 1.7.  The broken line corresponds to an unabsorbed $\Gamma$ =1.9
power-law.
}
\end{figure*}

\section{SWIRE Lockman survey}

The Spitzer SWIRE survey is described by Lonsdale et al (2003, 2004).  Source associations,
spectral energy distributions, photometric redshifts and infrared galaxy populations are
discussed by Rowan-Robinson et al (2005, 2008).  

There are 8563 SWIRE sources from the SWIRE photometric catalogue (Rowan-Robinson 
et al 2008) in the CLASXS-SWIRE area.
Of these,  233 are associated with the 401 extragalactic CLASXS X-ray sources, using
a 1.5" search radius between the respective optical positions.  
To see if we can find SWIRE associations for the remaining X-ray sources we have
examined the SWIRE sources that do not make it into the SWIRE photometric catalogue,
either because they are optically blank or do not meet the requirement of a total of two photometric
bands out of UgriZ, 3.6, 4.5 $\mu$m, of which at least one must be gri.  A total of 2192 further SWIRE sources,
satisfying either S(3.6) $>$ 10 $\mu$Jy and S(4.5) $>$ 8 $\mu$Jy, or S(24) $>$ 100 $\mu$Jy, fall
in the CLASXS-SWIRE area and 73 of these gave X-ray associations. 
So a total of 306 out of 401 extragalactic X-ray sources in CLASXS have SWIRE associations
(76$\%$).  Although Spitzer 
detects a much larger population of galaxies than Chandra, the SWIRE 
catalogue still contains only 75$\%$ of the X-ray sources.  The 96 non-SWIRE X-ray sources are
shown in Fig 2R.  They are typical of higher redshift X-ray sources and the non-detection by SWIRE 
does not imply unusual infrared properties.

The number of SWIRE sources which are either Type 1 QSOs or have significant
AGN dust tori components (contributing $\ge 20 \%$ of the 8 or 24$\mu$m emission) consists of 156 X-ray sources 
and 275 non-X-ray sources.  So only 36$\%$ of the galaxies recognized as AGN by SWIRE, 
either through a QSO template in the optical or through the presence of an AGN dust torus,
are detected in CLASXS.  Thus both surveys are needed to fully characterize the AGN population.
SWIRE tends to miss some of the weaker dust tori, whereas CLASXSmisses X-ray absorbed
or lower X-ray luminosity sources which SWIRE can detect through their dust tori.  
Donley et al (2005) first demonstrated the existence of AGN detected in the mid infrared by Spitzer 
through their dust tori, but undetected in deep Chandra integrations.  Comparing a deep Chandra integration
in the Groth strip with deep Spitzer data, Barmby et al (2006) found that 90$\%$ of their 150
X-ray sources were detected by IRAC and two thirds by MIPS.

If we regard 24 $\mu$m detection as essential for characterization of dust emission, then
200 of the 401 extragalactic CLASXS X-ray sources (50$\%$) are SWIRE 24 $\mu$m sources.
Figure 6 shows the SWIRE 24 $\mu$m flux versus the hard X-ray flux for these sources.  
There is no detailed correlation and a
very wide range of (S24$/$Sxh) ratio from $\sim$0.004  to 1.0 (mJy$/10^{-14}$ erg s$^{-1}$ cm$^{-2}$) .  The lowest 
values are candidates for very weak, low covering factor, dust tori, and high values can indicate 
dust tori with very high covering factors, sources with X-ray absorption in the hard band which
may be Compton thick sources, or possibly cases where X-ray emission is due to a starburst
rather than an AGN.  All these possibilities are discussed below.

\begin{figure*}
\epsfig{file=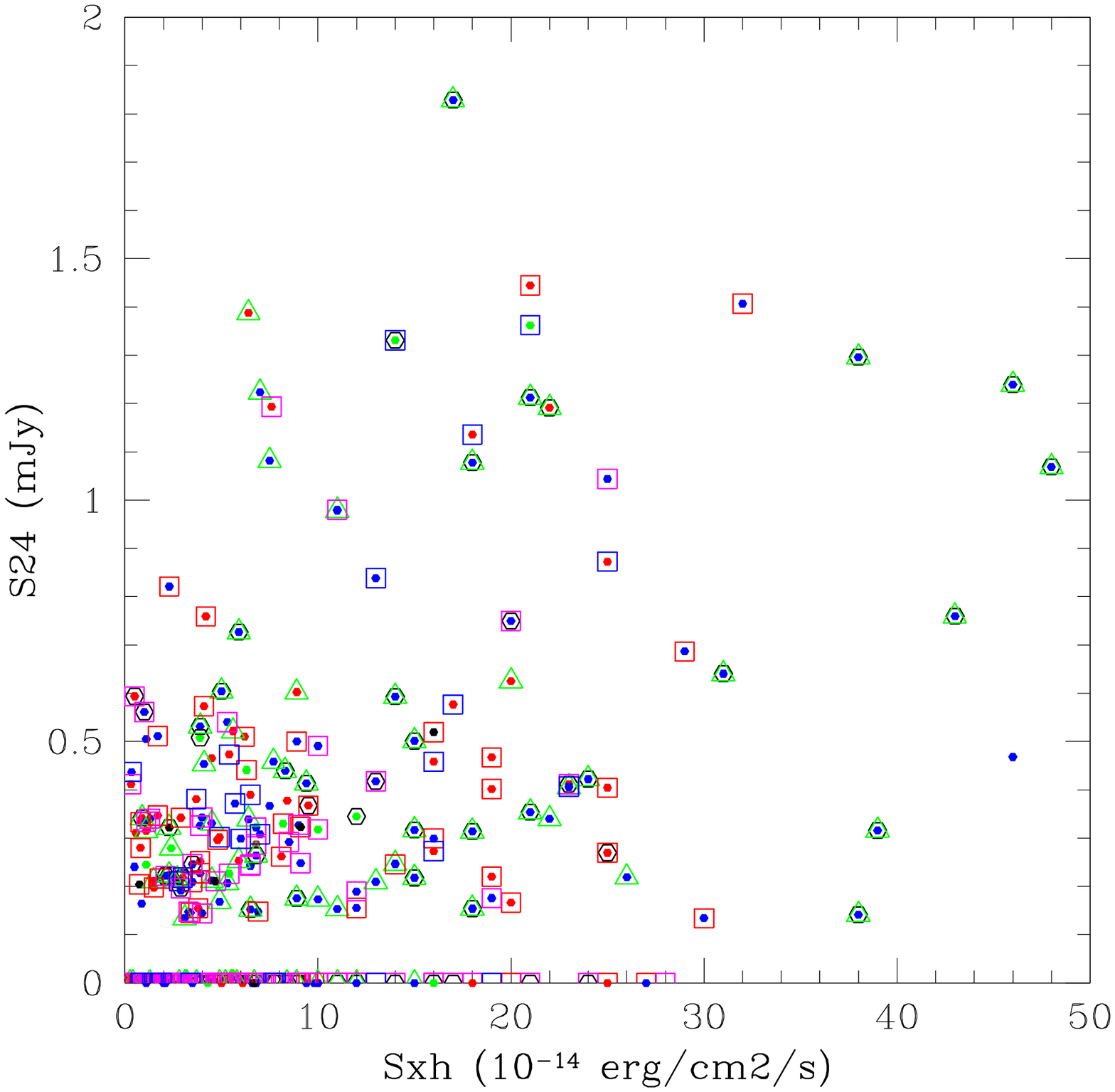,angle=0,width=14cm}
\caption{24 $\mu$ flux (mJy) versus hard X-ray flux ($10^{-14} erg s^{-1} cm^{-2}$) .
}
\end{figure*}

\subsection{Dust covering factor and Type 2 fraction}
Our ability to detect AGN dust tori at mid infrared wavelengths gives an excellent insight into the median dust covering
factor for AGN.  We fit infrared templates only if there are at least two infrared bands with infrared excess relative to the
starlight or QSO fit to optical data, and require that one of these bands must be 8 or 24 $\mu$m.  Because AGN dust
tori emit down to rest-frame wavelengths of 1 $\mu$m, in many cases dust tori are detected at all {\it Spitzer}
wavelengths from 3.6-24 $\mu$m (see Figs 11 and 12 below).  Properties of infrared detected AGN dust tori are
summarized in Table 1(b), binned by $L_{tor}$, and of the combined SWIRE-CLASXS AGN sample in Table 1(c). 

Figure 7L shows $L_{tor}$ versus $L_{Xh,c}$.  Objects to the left of the locus of equal luminosity
are either suffering from X-ray absorption or have an X-ray bolometric correction greater than the assumed
value of 27.  We define the covering factor by the dust tori to be the ratio of the luminosity in the dust torus to
the total bolometric luminosity of the AGN.  From Fig 7L, the covering factor appears to range from 0.01 to 1.

\begin{figure*}
\epsfig{file=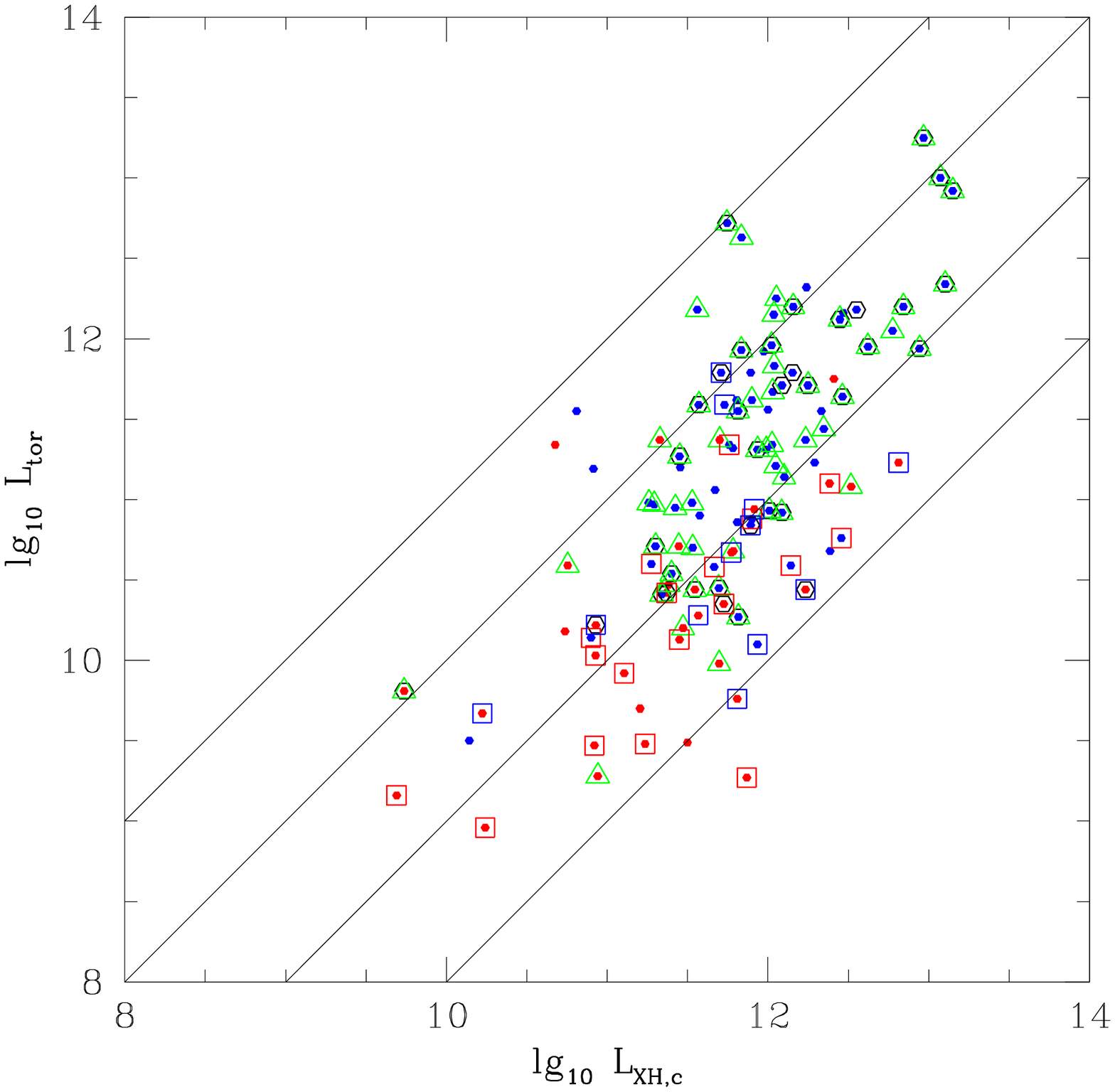,angle=0,width=7cm}
\epsfig{file=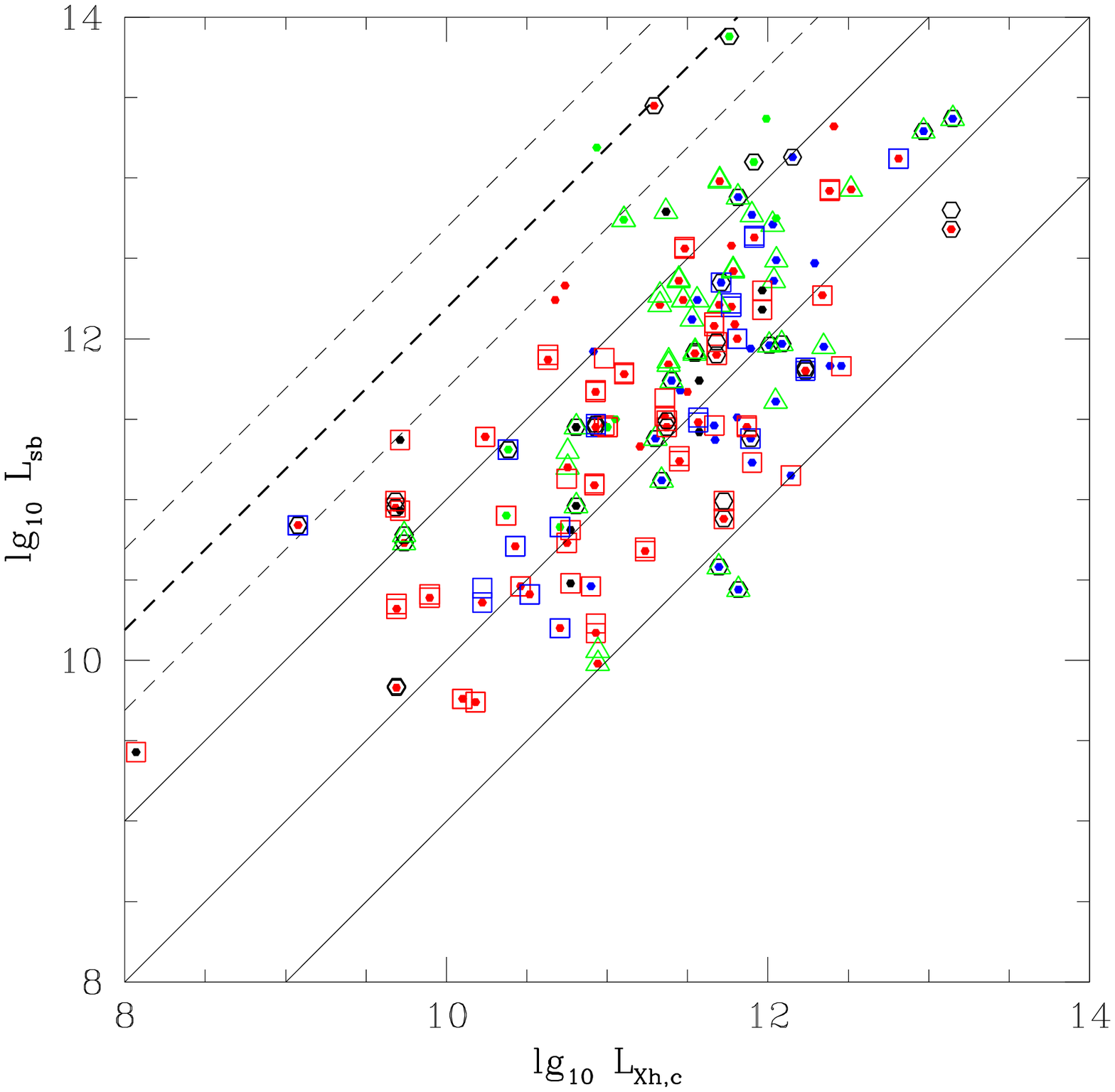,angle=0,width=7cm}
\caption{LH: AGN dust torus luminosity, $L_{tor}$, versus bolometric X-ray luminosity, $L_{Xh,c}$.
RH: Starburst luminosity, $L_{sb}$, versus bolometric X-ray luminosity, $L_{Xh,c}$.  The broken lines
denote  the Ranalli et al (2003) relation for X-ray emission from starbursts.
}
\end{figure*}

We define $L_{bh}$, as the total (X-ray to 3 $\mu$m) bolometric luminosity of the AGN, excluding starburst 
or dust torus emission.  For Type 1 QSOs, we 
assume a UV bolometric 
correction to the 0.1-3 $\mu$m luminosity, $L_{opt}$, of 2 (Rowan-Robinson et al 2008).  Otherwise we estimate
$L_{bh}$ from the X-ray luminosity, $L_{Xh,c}$, using the bolometric correction of 27 derived for this sample 
in section 2, bearing in mind that this will only be accurate on average.  In the light of the discussion of Fig 5 above,
we estimate $L_{bh}$ from X-ray luminosity even for Type 1 QSOs (i) if N(H) $> 10^{23} cm^{-2}$, (ii) if they are flagged 
as optically extended.

Figure 8 shows $L_{tor}/L_{bh}$ versus $L_X$.  We have indicated by vertical lines the luminosities below
which X-ray sources may be due to starbursts, and above which they are almost certainly QSOs.  The
horizontal line at $L_{tor}/L_{bh}$ = 1 corresponds to a dust torus covering factor of 100$\%$.
Objects above this line are candidates for being Compton thick AGN, though they could simply be 
cases of a higher than average X-ray bolometric correction if they are not Type 1 objects.  

The horizontal
broken line corresponds to the mean dust torus covering factor of 40 $\%$ derived by
Rowan-Robinson et al (2008) for a large sample of SWIRE QSOs with $L_{tor} > 10^{11.5} L_{\odot}$, from $<L_{tor}/L_{opt}>$.  
We can also estimate the mean covering factor for the present sample, from $<L_{tor}/L_{bh}>$, 
using luminous X-ray QSOs ($log_{10} L_{Xh} > 44.5$).  We find a slightly lower mean
covering factor of 35$\%$, but this is based on only 9 QSOs and is not inconsistent with the 40 $\%$ derived
by Rowan-Robinson et al (2008) using 796 QSOs.  The fraction of QSOs, as determined from optical template
fitting, at $log_{10} L_{Xh} > 44.5$ is 25$/$46 (54$\%$) again consistent with a 40 $\%$ covering factor. 

Objects with
very low values of $L_{tor}/L_{bh}$ are candidate for having very weak or non-existent tori and are discussed in
section 5 below.

\begin{figure*}
\epsfig{file=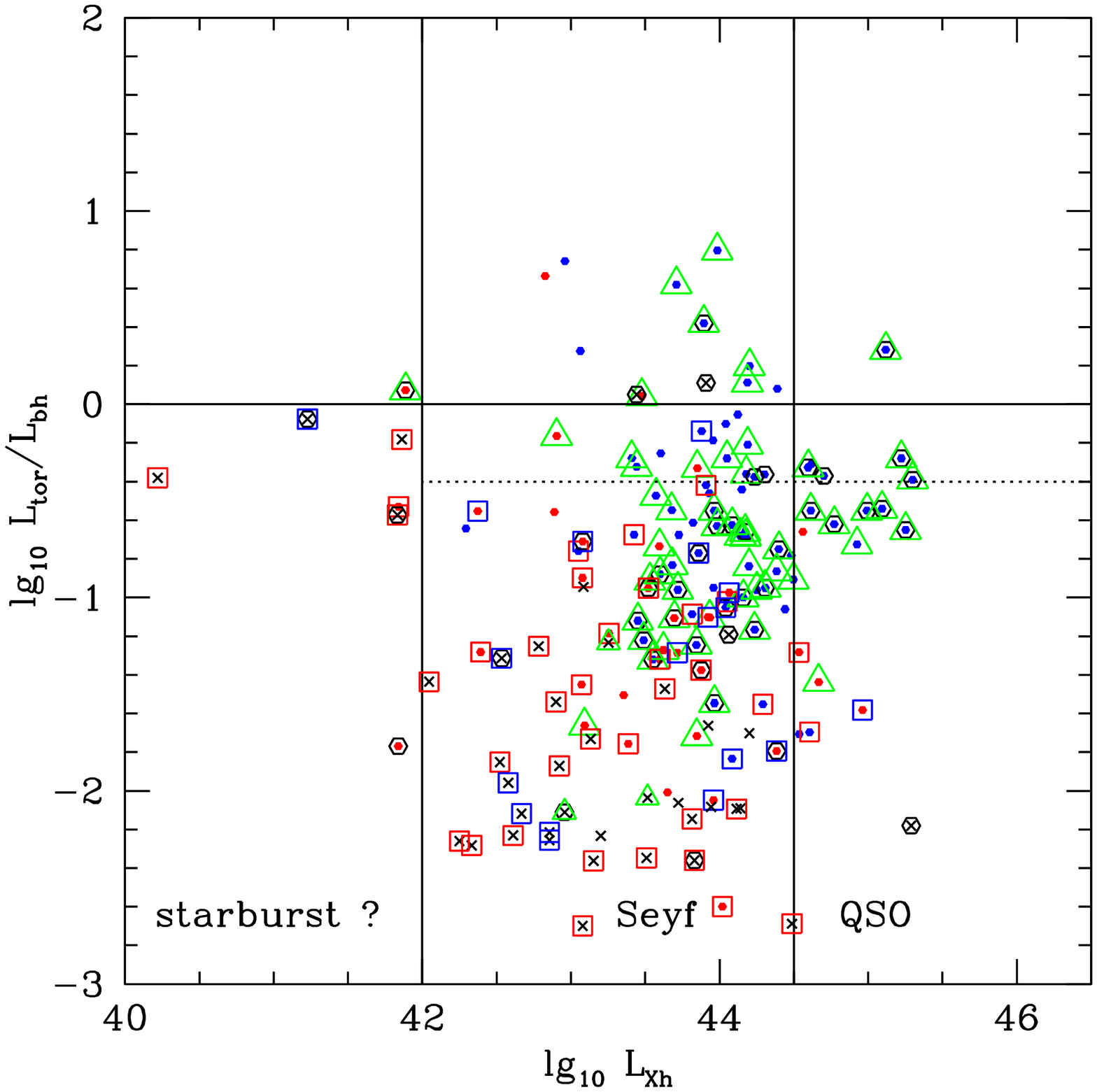,angle=0,width=14cm}
\caption{ Ratio of torus luminosity to bolometric optical-X-ray luminosity
versus X-ray luminosity.  Crosses denote upper limits for AGN with no dust torus detected by SWIRE.
}
\end{figure*}

For sources with no X-ray detection we show $L_{tor}/L_{opt}$ versus $L_{tor}$ in Fig 9.  We use $L_{tor}$
as the x-axis because in all cases this is an, albeit indirect, measure of the AGN luminosity.  We have plotted this
for SWIRE AGN dust tori with no hard X-ray detection (Fig 9L) , and the hard X-ray AGN (Fig 9R), separately.  
For X-ray sources the distribution is broadly equivalent to Fig 8, except that lines of constant $L_{opt}$ now have
slope +1 (we have indicated with a broken line the approximate boundary between galaxies and QSOs).
AGN which have optical QSO templates and$/$or broad lines show a trend in which the covering factor
($L_{tor}/L_{opt}$) decreases as $L_{tor}$ decreases.  Such a trend can also be seen in Fig 8.
For AGN dust tori associated with galaxies (the overwhelming majority of the non-X-ray AGN),
the sources may either be Type 2 objects, in which the QSO is shrouded by dust, or weaker Seyferts 
where we do not see the optical AGN against the light of the host galaxy.   $L_{tor}/L_{opt}$
will underestimate the true covering factor for Type 1 Seyferts. 
However for objects with $L_{tor} > 2 L_{opt}$ we can be confident 
they are Type 2 objects, since even for an almost 100$\%$ covering factor the QSO would outshine the galaxy if
viewed pole-on.  $L_{tor}/2 L_{opt}$ clearly overestimates the covering factor for these objects.
Despite these limitations it is of interest to see how the median value of  $L_{tor} / 2 L_{opt}$ for the combined
infrared and X-ray sample varies with torus luminosity and we have indicated this in the last two lines of Table 1 (c). 
We find that  $L_{tor}/2 L_{opt}$ decrease as $L_{tor}$ is reduced, from 0.4 in the highest luminosity bin, consistent 
with the result of Rowan-Robinson et al (2008), to 0.08 in the lowest luminosity bin (this trend is indicated by
the dotted line in Fig 9).   However the lowest luminosity dust tori are probably simply Seyferts, AGN with optical luminosities
much less than that of their parent galaxy, so  $L_{tor}/2 L_{opt}$ severely underestimates the covering dust factor.
In Table 1(a) we have shown the median value of $lg(L_{tor}/L_{bh})$ for the X-ray sample.
Our data are consistent with a model in which the median covering factor at first decreases from high to moderate luminosity,
then increases towards low luminosities.    This would run counter to several
claims in the literature, which suggest that the covering factor increases monotonically towards lower luminosities. 

Hasinger (2008) showed that the ratio of narrow-line 
X-ray AGN to broad-line AGN increases with decreasing hard X-ray luminosity, using nine X-ray surveys of which
CLASXS is one.  A similar trend is found also by e.g. Ueda et al.
(2003), La Franca et al. (2005), Gilli et al. (2007), and Fiore et al.
(2008).   However all these studies use the fraction of absorbed AGN
(from X-rays) or broad line AGN (from optical spectroscopy) which give information
on the total line-of-sight absorption, which may include galaxy-scale absorption (for Seyfert 1.8 and1.9, Maoilino 
and Rieke 1995) or dust-free ionized gas (Comastri et al 2001, Nandra et al 2003).
In Table 1 we show the corresponding numbers of narrow-line and broad-line AGN in our sample, and their
ratio, which does indeed show this effect. However there are some problems with immediately interpreting this
as an increase of covering factor with declining luminosity.  The presence of broad lines, or of an optical QSO continuum,
 is not a reliable indicator that an object is face-on.  Several of the objects that we believe have to be Type 2,
Compton-thick AGN, on the basis of $L_{tor}/L_{bh}$ or $L_{tor}/L_{opt}$, have broad-line spectra and$/$or an
optical QSO continuum (see section 4 and Table 2).  Polletta et al (2006)'s object SWIRE-J104409.95+585224.8 
has $L_{tor} \sim 100 L_{opt}$, yet its optical spectrum
is that of an unreddened broad-line QSO, clearly because it is seen in scattered light in the optical.  A second 
problem with arguments based on broad$/$narrow line ratios is that they are based on a small fraction of the
X-ray AGN sample (22$\%$ of the AGN with $L_{Xh} < 10^{43.5}$ erg s$^{-1}$ in Table 1(a)). 
If we look more closely at the lowest luminosity bin, where 5 out of 7 AGN are narrow-line, it turns out that 4 of these
are low absorption AGN (the 5th has $log_{10}$N(H) = 22.38) and for 3 of them their optical SEDs are fitted with QSO
templates.  So these 5 objects do not on their own constitute a strong case for a high covering factor at low luminosities. 

To investigate this further we have divided all CLASXS objects spectroscopically classified as broad-line or narrow-line
into unabsorbed ($log_{10} N(H) < 22$), absorbed ($22<log_{10} N(H) < 23$) and strongly absorbed ($log_{10} N(H) >23$).
In the unified model for AGN we might have expected that all broad-line objects would be
unabsorbed and that all narrow-line objects would show at least some absorption. 
The actual numbers of broad-line objects in the 3 classes are 53, 31 and 5 respectively, and the numbers of narrow-line
objects are 15, 15 and 7 respectively.  So the broad$/$narrow-line distinction is not a very good indicator of
X-ray absorption.  The broad-line objects with strong X-ray absorption presumably may imply that the broad lines are scattered round the
X-ray absorbing cloud or clouds. Alternatively there may be a link with BAL QSOs which are generally not very strongly absorbed in the
optical but are characterized by strong X-ray absorption (Gallagher et al 2006, Fan et al 2009), due to a dust-free outflowing medium.
 To explain the narrow-line unabsorbed 
objects we must presumably invoke excess soft X-ray emission masking the absorption.

Maiolino et al (2007) have presented evidence of increasing Type 2 fraction as luminosity is reduced, based on
the ratio of the 6.7 $\mu$m  to the 5100 A continuum, which should be equivalent to our $L_{tor}/L_{opt}$ ratio.  
However this is not supported by the much larger and perhaps more representative sample of our Fig 9.  
Treister et al (2008) have also argued for
an increasing dust covering factor towards lower luminosities by considering the ratio of 24 $\mu$m luminosity to
bolometric luminosity.  We believe it is crucial to separate the torus and starburst 
contributions to the 24 $\mu$m emission. 

Some authors (eg La Franca et al (2005), Hasinger 2008, Treister et al 2008) have suggested that the fraction 
of obscured AGN increases significantly with redshift.  Because higher luminosity AGN tend to be at higher redshifts
in flux-limited samples like ours, this could account for part of the increase in covering factor seen for the highest
luminosity bin in Table 1.   Our sample is probably not large enough to disentangle luminosity and redshift dependence.

\begin{table*}
\caption{Properties of X-ray and infrared AGN in SWIRE-CLASXS sample, as a function of luminosity}
\begin{tabular}{llllll}
property & $L_{bh} < 10^{10.35}$ & $10^{10.35} < L_{bh} < 10^{11.35}$ & $10^{11.35} < L_{bh} < 10^{12.35}$ &  $L_{bh} > 10^{12.35}$ & {\bf total} \\
& ($L_{\odot}$) &&&&\\
{\bf (a) X-ray detected AGN} & & & & &\\
 $L_{Xh}$ (erg $s^{-1}$) & $ <10^{42.5}$ & $10^{42.5} - 10^{43.5}$ & $10^{43.5} - 10^{44.5}$ & $> 10^{44.5}$ &\\
 &&&&&\\
total number & 40 & 117 & 198 & 46 & {\bf 401}\\
X-ray starbursts & 4 & 5 & 1 & 0 &  {\bf 10}\\
optical QSO template & 9 & 21 & 51 & 25 &  {\bf 106}\\
N(H) $> 10^{24} cm^{-2}$ & 0 & 0 & 0 & 1 &  {\bf 1}\\
$10^{23}<$N(H)$<10^{24}$ & 0 & 8 & 61 & 23 & {\bf 92}\\
N(H) $< 10^{22} cm^{-2}$ & 34 & 63 & 53 & 9 & {\bf 159}\\
$L_{tor} > L_{bh}$ & 0 & 4 & 8 & 1 &  {\bf 13}\\
no. narrow-line AGN & 5 & 15 & 14 & 3 & {\bf 37}\\
no. broad-line AGN & 2 & 12 & 57 & 18 & {\bf 89}\\
ratio(nl AGN$/$bl AGN) & 2.5 & 1.2 & 0.25 & 0.17 & {\bf 0.42}\\
median lg($L_{tor}/L_{bh}$) & -0.60 & -0.75 & -0.80 & -0.58 &\\
implied covering factor & 25$\%$ &  18$\%$ & 16$\%$ & 26$\%$ & \\
&&&&&\\
{\bf (b) infrared detected AGN dust tori} & & & & &\\
{\bf  (no X)} $L_{tor} (L_{\odot})$ & $ <10^{9.95}$ & $10^{9.95} - 10^{10.95}$ & $10^{10.95} - 10^{11.95}$ & $> 10^{11.95}$ &\\
&&&&&\\
total number & 56 & 175 & 18 & 8 &  {\bf 257}\\
optical QSO template & 0 & 0 & 0 & 7 &  {\bf 7}\\
N(H) $> 10^{24} cm^{-2}$ & $\ge$0 & $\ge$92 & 16 & 8 & {\bf $\ge$116}\\
N(H) $> 10^{23} cm^{-2}$ & 47 & 174& 18 & 8 &{\bf  247}\\
$L_{tor} > L_{opt,c}$ & 1 & 12 & 3 & 1 &  {\bf 17}\\
&&&&&\\
{\bf (c) combined sample} & 96 & 292 & 216 & 54 &  {\bf 658}\\
(excluding starbursts) &&&&&\\
no. optical QSO template & $\ge$9 & $\ge$21 & $\ge$51 & 32 &  {\bf $\ge$114}\\
$\%$ QSO & $\ge 8 \%$ & $\ge 7 \%$ & $\ge 24 \%$ &  60$\%$ &\\
no. Compton-thick & $\ge$0 & $\ge$96 & 24 & 10 &  {\bf $\ge$130}\\
$\%$ Compton-thick & $\ge 0\%$ & $\ge 33\%$ & $\ge11 \%$ & 18$\%$ &  {\bf $\ge$ 20$\%$}\\
no. N(H) $> 10^{23} cm^{-2}$ & 47 & 182& 79 & 32&{\bf  247}\\
$\%$ N(H) $> 10^{23} cm^{-2}$ & 50 & 63& 39 & 58&{\bf  41}\\
median lg($L_{tor}/2L_{opt})$ & -1.1& -0.55 & -0.7 & -0.4 &\\
implied covering factor & 8$\%$ & 28$\%$ & 20$\%$ & 40$\%$& \\
\end{tabular}
\end{table*}

\begin{figure*}
\epsfig{file=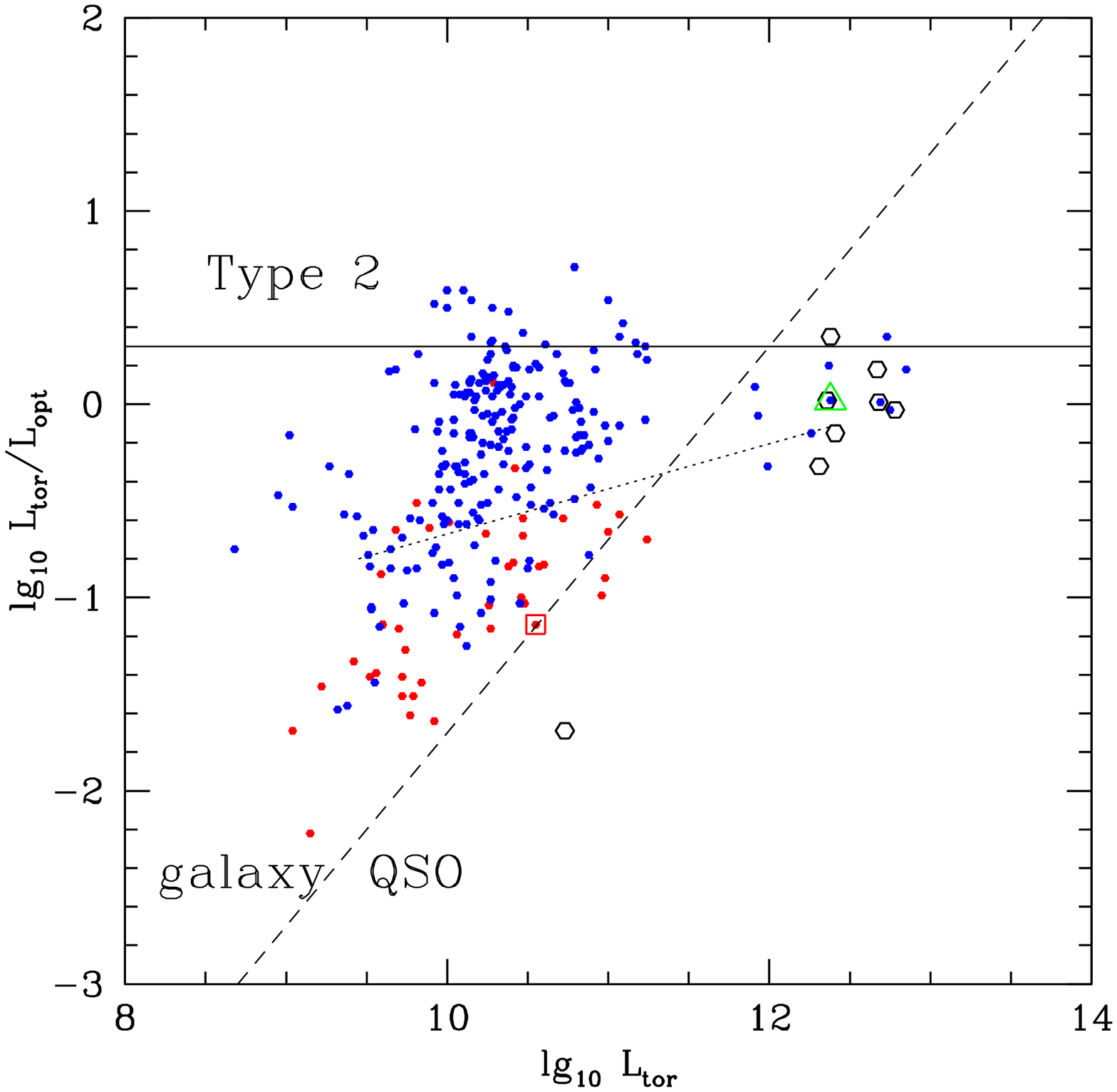,angle=0,width=7cm}
\epsfig{file=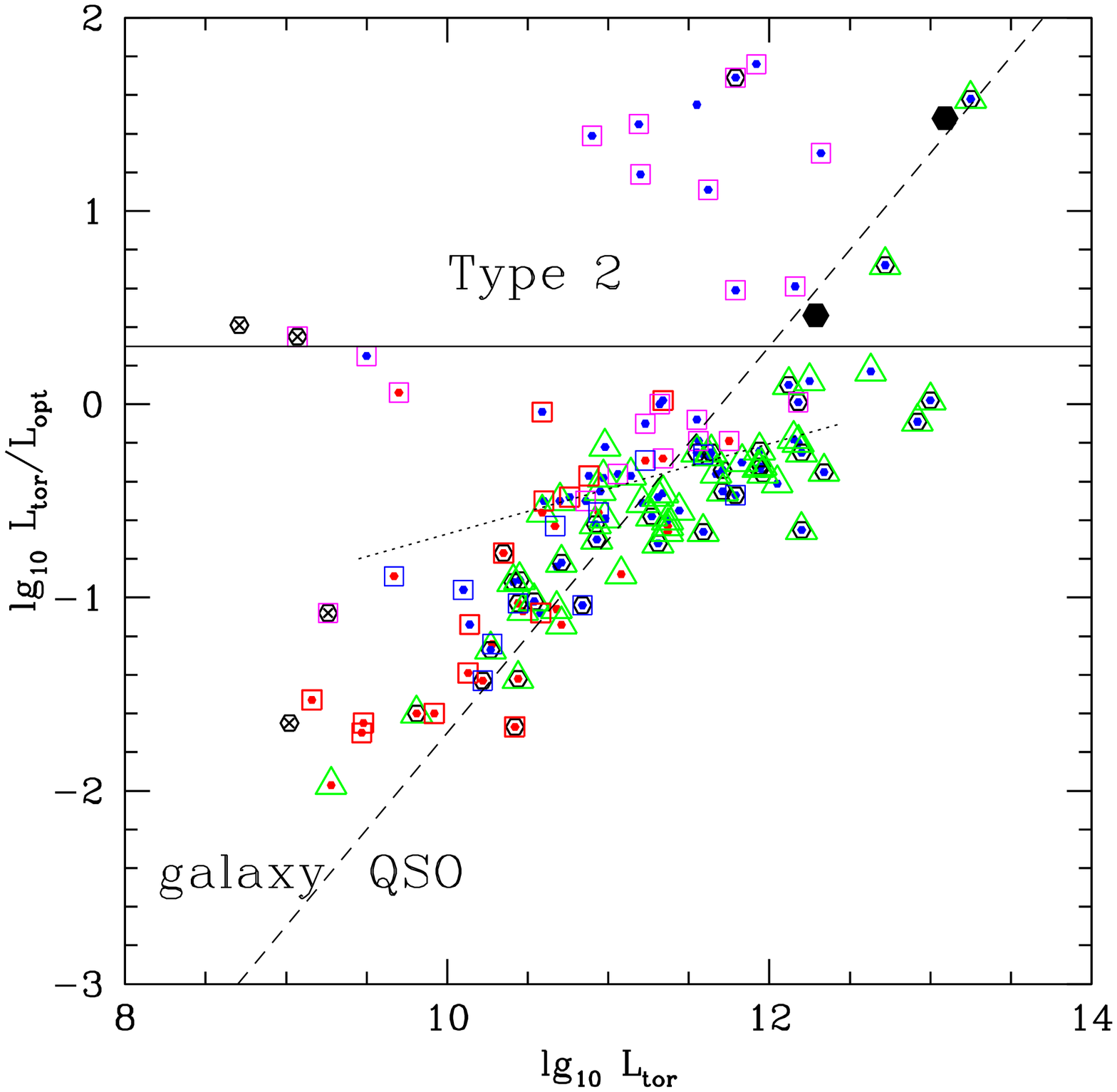,angle=0,width=7cm}
\caption{Ratio of torus luminosity to optical luminosity versus torus luminosity, for: (L) infrared-selected
AGN (SWIRE sources with optical QSO templates, or presence of AGN dust tori) and no hard X-ray detection,
(R) Hard X-ray AGN sources with optical QSO templates, or presence of AGN dust tori .  
The large filled circles are the two Compton-thick QSOs studied by Polletta et al (2006).
The broken line indicates the optical luminosity above which objects are likely to be QSOs.
The dotted line indicates the trend of the median $L_{tor}/L_{opt}$ for the whole SWIRE-CLASXS sample
(last line of Table 1).
}
\end{figure*}

\subsection{Fraction of Compton-thick AGN}
Figure 10 shows $L_{tor}$ versus redshift (LH) and $L_{Xh}$ versus redshift (RH), illustrating the much
greater sensitivity of SWIRE to lower-luminosity, lower-z, dust tori, while Chandra sees deeper in the 
high-luminosity sources.  It is apparent that many of the lower luminosity dust tori would by easily detectable
in X-rays if they had similar properties to the X-ray detected objects.  To quantify this we have estimated the
unabsorbed X-ray luminosity each of these sources would have with a hard X-ray flux at the limit of the
CLASXS survey ($3.10^{-15} erg s^{-1} cm^{-2}$), and then used the ansatz

$L_{bh} \sim 27 L_{Xh} \sim 2 L_{opt} \sim 2.5 L_{tor}$         (1)

appropriate to a dust torus with the average optical and X-ray bolometric corrections, and average
dust covering factor of 40$\%$, to estimate how much X-ray absorption would be needed to push
the X-ray flux down to this limit.  116 out 246 infrared AGN dust tori not detected in X-rays were found to need
a minimum hydrogen column N(H) of $\sim 10^{24}$ cm$^{-2}$, corresponding to Compton-thick objects.
Table 1 summarizes some properties of the X-ray and non-X-ray sources in the SWIRE-CLASXS sample,
as a function of luminosity class.
Our conclusion is that $\ge 20\%$ of the sample is Compton thick, and that $\ge 52\%$ have N(H) $> 10^{23} cm^{-2}$,
but we are not able to demonstrate
any clear dependence of these percentages on X-ray or torus luminosity.   $\sim 25\%$ of the sample have
N(H) $< 10^{22} cm^{-2}$ and are unabsorbed.  Again there is little dependence on AGN luminosity except
perhaps that the the proportion of unabsorbed sources is higher in  the lowest bin of X-ray or torus
luminosity. 
Our results therefore favour the X-ray background model of Treister and Urry (2005), who assume a 25 $\%$ fraction
of unabsorbed sources independently of luminosity and of redshift.
However we could not rule out a
model in which the Compton-thick fraction went from 20$\%$ at $L_{Xh} = 10^{44.5}$ erg s$^{-1}$ to 50$\%$
at $L_{Xh} = 10^{42.5}$ erg s$^{-1}$, and such a behaviour might contribute to a viable model of the hard X-ray background.

\begin{figure*}
\epsfig{file=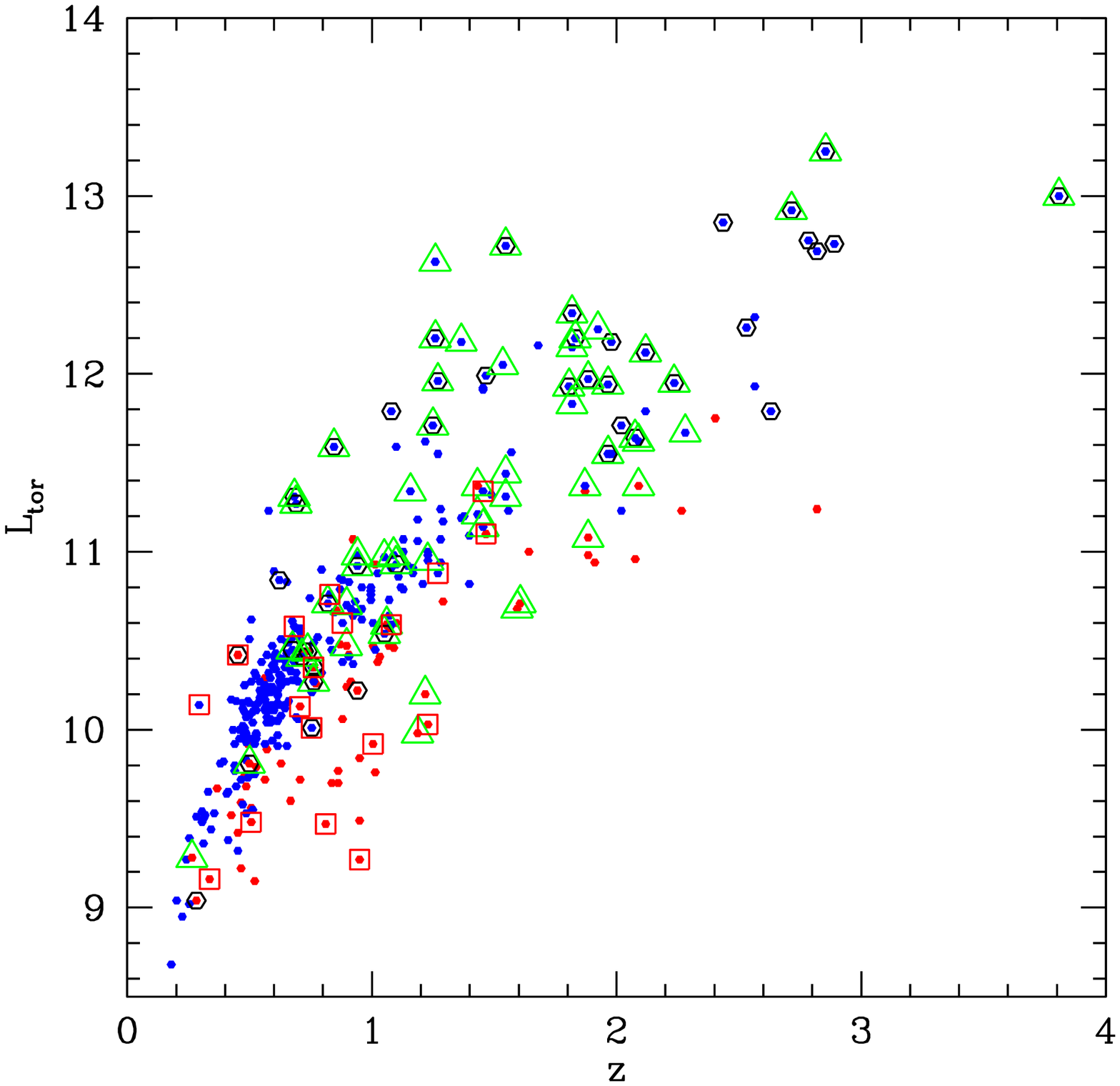,angle=0,width=7cm}
\epsfig{file=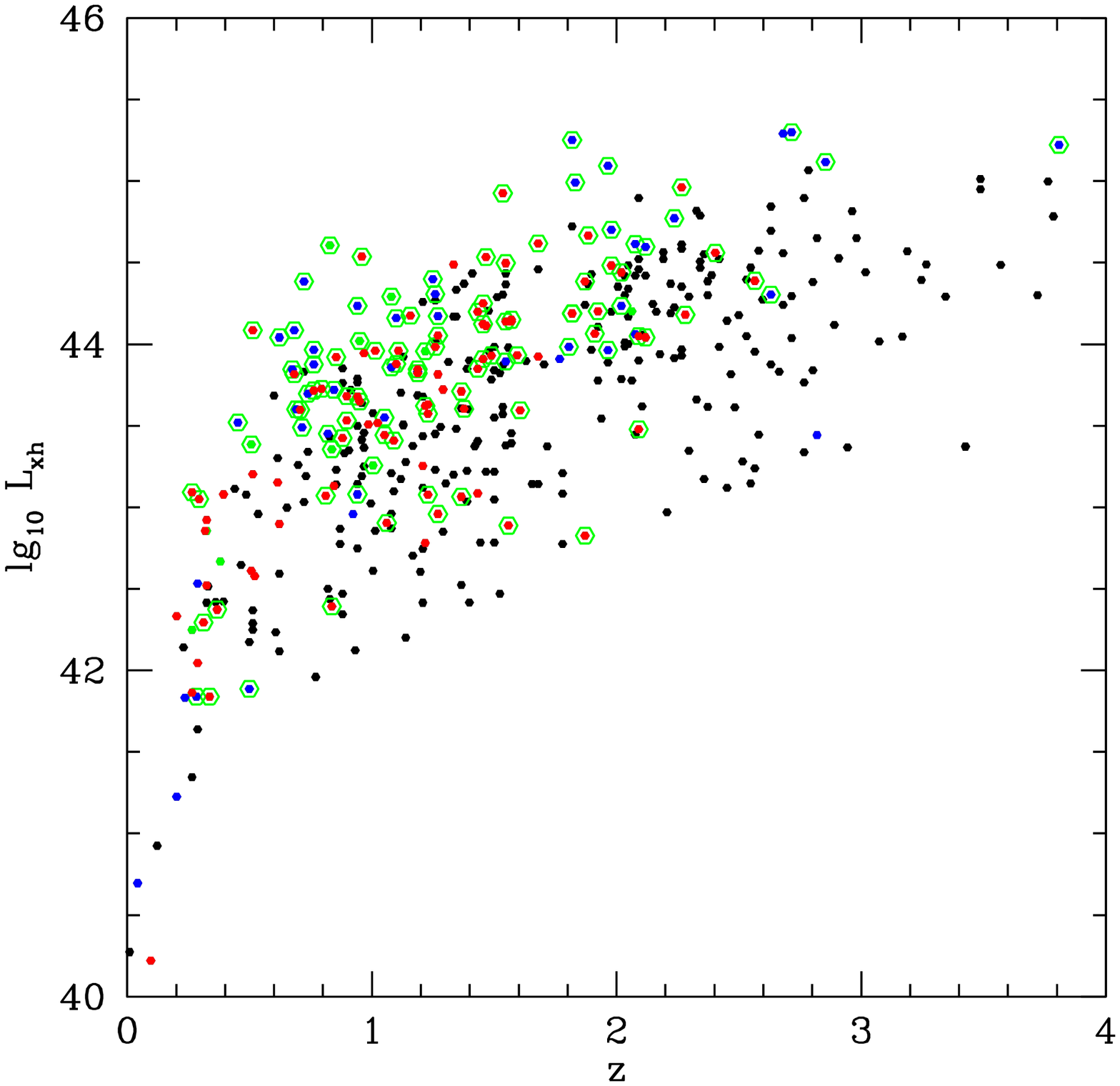,angle=0,width=7cm}
\caption{LH: Torus luminosity versus redshift.  Filled red circles: starburst ir SEDs, filled blue circles: 
dost torus dominated; green open triangle: broad-line; blue open square: narow-line; red open square: 
galaxy spectra; black open circle: optical QSO.
RH: X-ray luminosity versus redshift.  Sources with infrared dust tori: open green circle; and with optical QSO (filled blue circle); 
with E-Sab (filled green circle); with Sbc-sb (filled red circle).  Other X-ray galaxies: filled black circle.
}
\end{figure*}

\section{Modelling the SEDs of Compton thick AGN}

The optical-infrared SED of the one Compton thick AGN identified from the ratio of hard to soft counts is shown in 
Fig 11L (object 1, Table 2a).  
It is a Type 1 QSO in its optical SED, and has 
  $L_{tor} > L_{opt}$,  but it does not have the vey high $L_{tor}/L_{opt}$ ratio of  the two best studied Compton-thick objects
of Polletta et al (2006). 
There may be more Compton-thick objects in the CLASXS X-ray sample because of the possibility of
excess soft X-ray emission from ionized gas (warm absorber model, Gierlinski and Done 2006), Compton 
scattered light (reflected from accretion disk, Nandra and Pounds 1994, Maloney and Reynolds 2000, Ross and Fabian 2005,
Nandra and Iwasawa 2007) or
inhomogenous covering (Winter et al 2008).  The latter is to be expected in a multi-cloud torus model. 

Fig 11 also shows 9 candidate Compton thick X-ray objects from the condition $L_{tor}/L_{bh} > 1$.  By definition all
have prominent dust tori.  Because of uncertainties in the bolometric correction factors used to calculate
$L_{bh}$, we can not be certain that any of these objects require enhanced X-ray absorption.  A further uncertainty
could be introduced by the possibility of X-ray variability in the interval between the {\it Chandra} and {\it Spitzer} observations.
The direct estimates
of N(H) are in the range $10^{21}$ - 4.$10^{23}$, and 3 are unabsorbed, but these may be underestimates because of the
effects of Compton scattering.  Objects 3, 5, 6, 10 in Table 2b and Fig 11, which have $L_{tor} >> L_{xh,c}$ and 
optical seds fitted with galaxy templates, are candidates for being obscured AGN of the type discussed by Ueda 
et al (2008), where the torus covering factor is close to 100$\%$.

A further 18 SWIRE objects not detected in X rays have $L_{tor} > L_{opt,c}$,
but 17 of these are galaxies with much lower redshifts and torus luminosities than the objects plotted in Fig 11
 and could be simply Type 2 QSOs.  However the estimate of the absorption required to explain their non-detection
in X-rays (see section 3) shows that most are probably Compton thick, and all have N(H) $> 10^{23.8}$ cm$^{-2}$.
The 18th object has a QSO optical SED, with photometric redshift 2.89,
and is very similar to the X-ray objects in Fig 11.  It has been included in Fig 11 (object 11). The detailed 
model of Fig 11 has  $L_{tor} < L_{opt,c}$ but the estimate of the absorption required to explain its non-detection
in X-rays shows that it is definitely Compton thick.  Properties of both types of
Compton thick candidates are given in Table 2.  Spectroscopic types in this and subsequent tables are: 
4 = broad-line QSO, 2,3 = narrow-line AGN, 1 = galaxy. Photometric redshifts are shown bracketed.

According to the AGN dust torus models of Efstathiou and
Rowan-Robinson (1995), a highly edge-on dust torus would have its emission peaking at a
rest-frame wavelength $\sim 30 \mu$m.  We have not seen any such objects but do not have enough 
sensitivity at 70 and 160 $\mu$m to readily detect such tori.

There are 73 X-ray sources in the SWIRE-CLASXS sample with z$>$2, and the number of Compton-thick
candidates among these is one directly from N(H), 4 from $L_{tor} > L_{opt,c}$, with perhaps a further 10 objects
missed from the sample due to strong X-ray absorption (section 3), corresponding to a Compton-thick fraction  15$/$83 or 19$\%$
(see also Table 1). 

\begin{figure*}
\epsfig{file=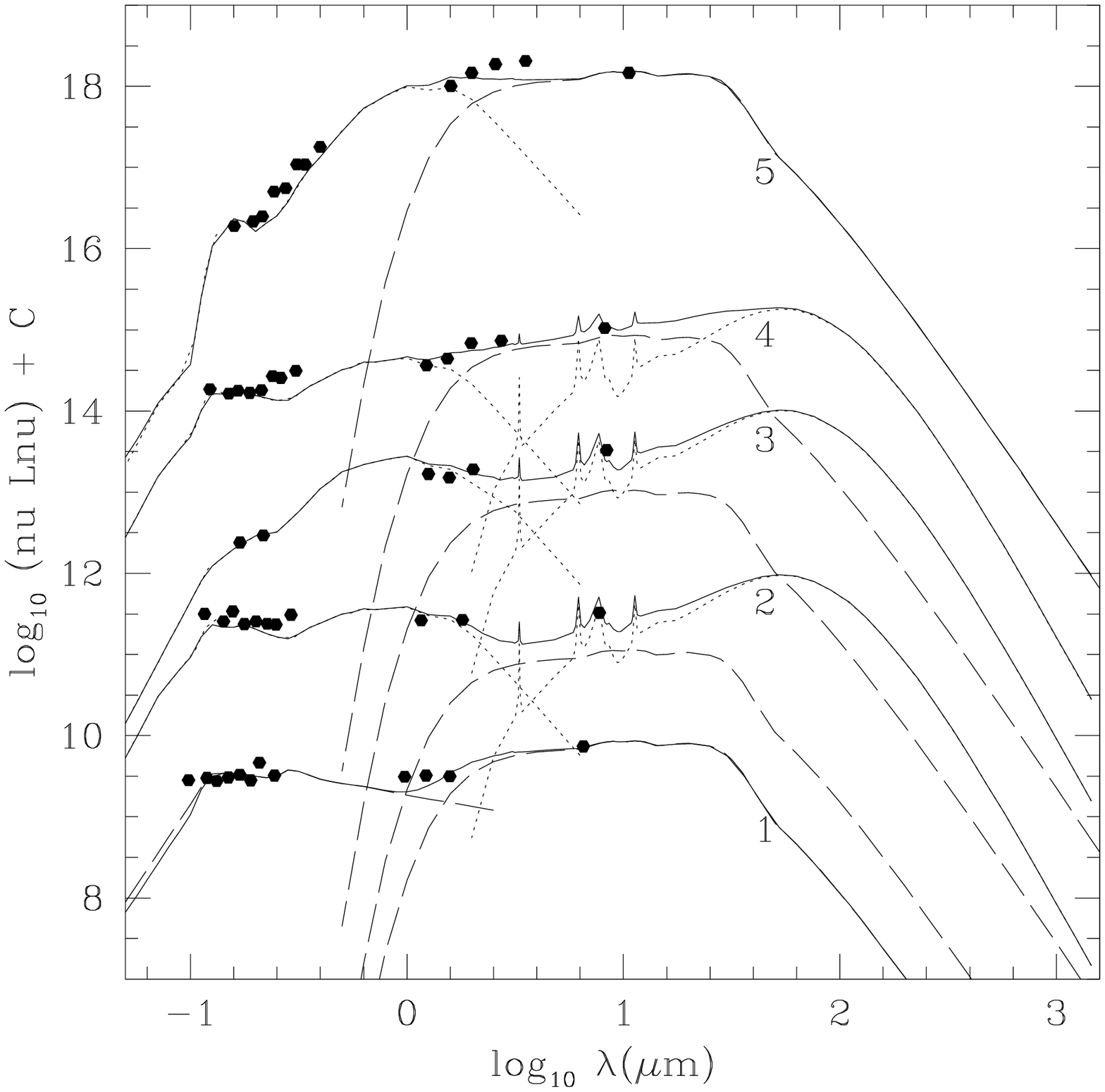,angle=0,width=7cm}
\epsfig{file=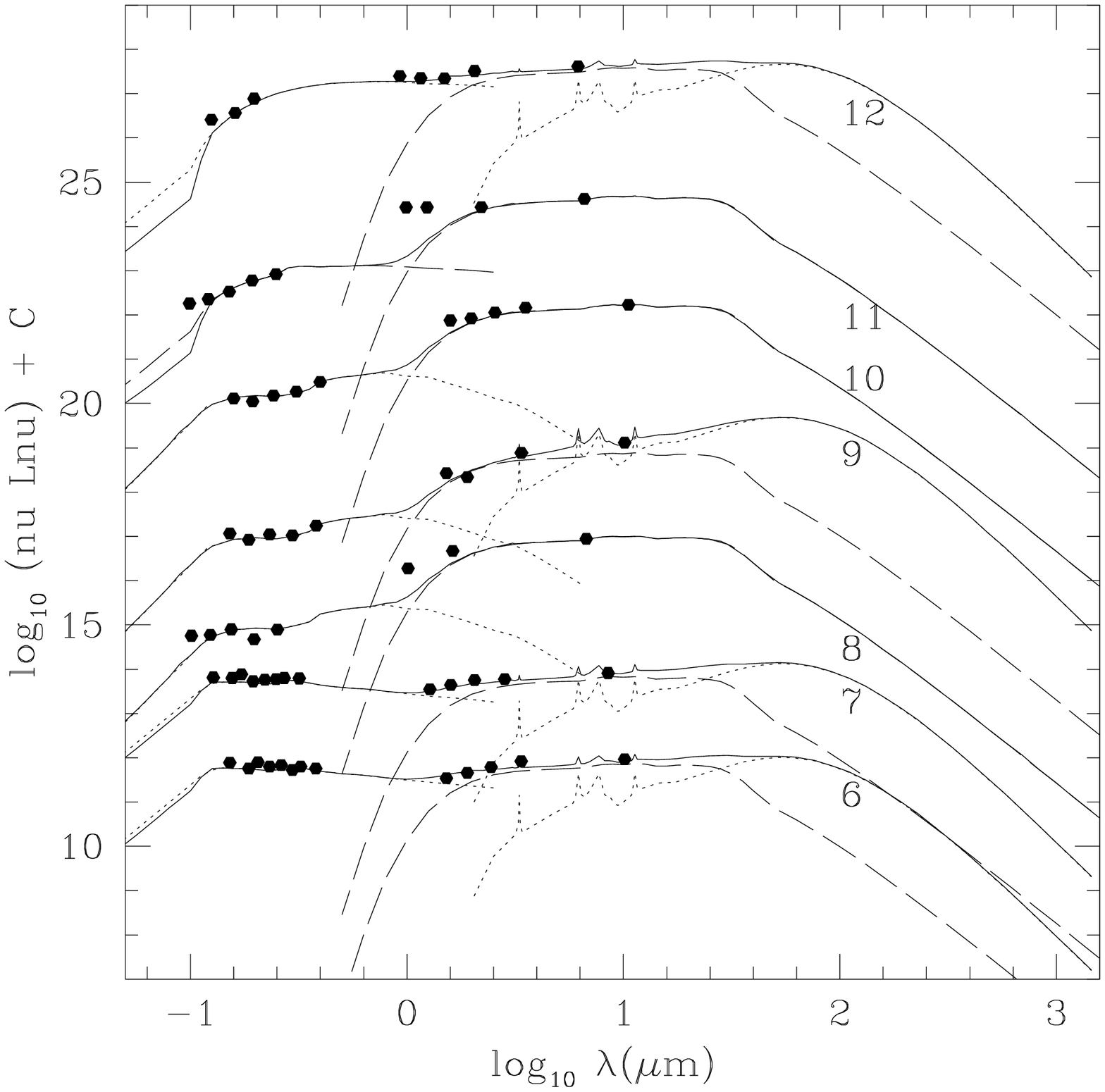,angle=0,width=7cm}
\caption{SEDs of candidate Compton thick AGN. Object 1 (see Table 2) has N(H) $> 10^{24}$ cm$^{-2}$ deduced from hardness ratio, 
objects 2-10 have  $L_{tor}/L_{bh} > 1$ and object 11 is not detected in X-rays but has $L_{tor} > L_{opt,c}$. 
}
\end{figure*}

\begin{table*}
\caption{Compton thick candidates: (a) $N(H) > 10 ^{24} cm^{-2}$, (b) $L_{tor} > L_{bh}$}
\begin{tabular}{llllllllllll}
object no. & RA & dec & z & spec. type & $L_{xhc}$ & $log_{10} N(H)$ & $L_{sb}$ & $L_{tor}$ & $L_{opt}$ & type & $A_V$\\
(a) &&&&&&&&&&&\\
1 & 158.55830 & 57.77818 & (2.68) &  & 13.14 & 24.04 &  & 12.27 & 12.17 & QSO & 0.2\\
(b) &&&&&&&&&&&\\
2 & 158.34232 & 57.54998 & 2.096 & 4 & 11.33 & 19.57 & 12.21 & 11.37 & 12.03 & sb & 0.1\\
3 & 158.68903 & 57.67629 & (1.87) & & 10.68 & 19.57 & 12.23 & 11.33 & 11.61 & Sbc & \\
4 & 158.33299 & 57.80566 & 1.928 & 4 & 12.05 & 21.13 & 12.48 & 12.24 & 12.12 & sb & 0.3\\
5 & 158.78539 & 57.96220 & 1.261 & 4 & 11.84 & 23.19 & & 12.45 & 12.65 & sb & 1.7\\
6 & 158.92015 & 57.84380 & 1.370 & 4 & 11.56 & 22.62 & 12.20 & 12.14 & 12.34 & sb \\
7 & 158.32928 & 57.85759 & 1.823 & 4 & 12.04 & 22.65 & 12.35 & 12.14 & 12.32 & sb \\
8 & 158.30424 & 57.56830 & (2.565) & & 12.24 & 22.85 & & 12.33 & 10.61 & Scd &\\
9 & 158.41524 & 57.63800 & (1.366) &  & 10.91 & 22.67 & 11.88 & 11.16 & 9.70 & Scd &\\
10 & 158.72195 & 57.56465 & 1.270 & & 10.81 & 19.57 & & 11.50 & 9.95 & Scd & \\
11 & 158.45297 & 57.70647& (2.63) & & 12.15 & 23.61 & & 12.02 & 11.07 & QSO & 0.7\\
12 & 158.66422 & 57.59955 & (2.89) & & & & 12.90 & 12.92 & 13.32 & QSO & 0.9\\
\end{tabular}
\end{table*}

\section{Weak tori AGN}

Figure 12L shows the SEDs of X-ray AGN which are candidates for having very weak tori
(properties given in Table 3).
For five objects (2, 4, 5, 6, 10) the photometric redshift  solution selected a highly reddened QSO, which
then gives values for $L_{bh}$ factors of 5-20 times $L_{Xh,c}$.  However they are all flagged as
optically extended objects and galaxy models seem much more plausible.  The ratios $L_{tor}/L_{Xh,c}$  become
0.2-0.3.
The remaining objects (1, 3, 7, 8, 9) appear to be Type 2 objects, with $L_{Xh,c} > L_{opt}$.
These show a range of covering factors from 1-3$\%$.

There were a further 3 objects which our automatic SED-fitting code gave strongly reddened QSO fits in the optical,
with Arp 220 or cirrus+starburst fits in the infrared, ie no dust torus.  However more careful modeling
of the SEDs showed that these were probably unreddened galaxy SEDs in the optical, with dust tori
and starbursts in the infrared, ie they are in fact Type 2 objects.
Thus there are no clear cases where the implied limit on the dust torus component
would correspond to a covering factor $< 1\%$.  
Our conclusion is that the dust covering factor around luminous AGN ranges from
1-100$\%$, with a median value of 40$\%$.
However dust covering factors of just a few percent
are not very different from naked quasars and these objects with very weak tori do not fit easily
into a unification model in which AGN properties depend only on viewing angle.

\begin{figure*}
\epsfig{file=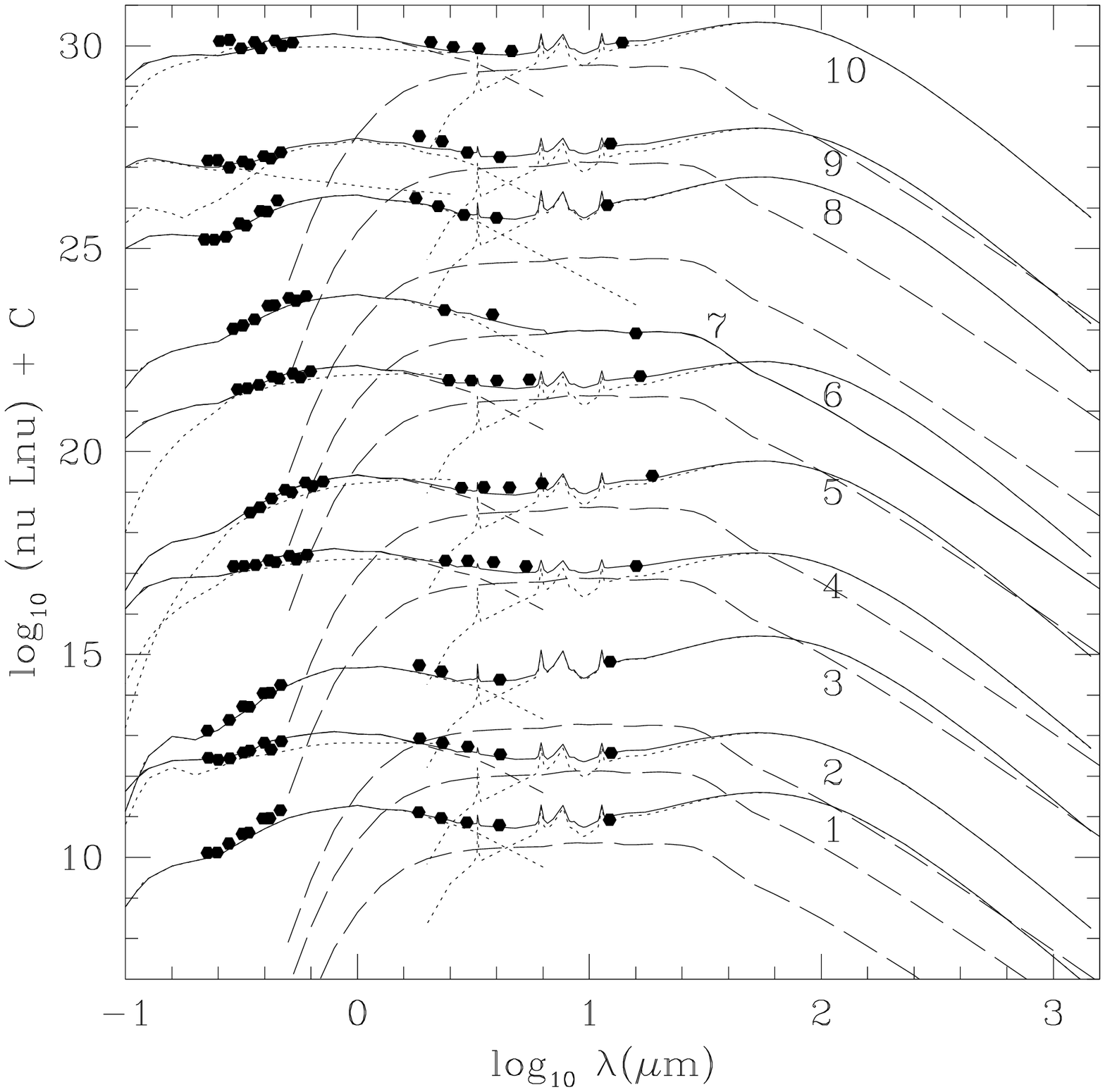,angle=0,width=7cm}
\epsfig{file=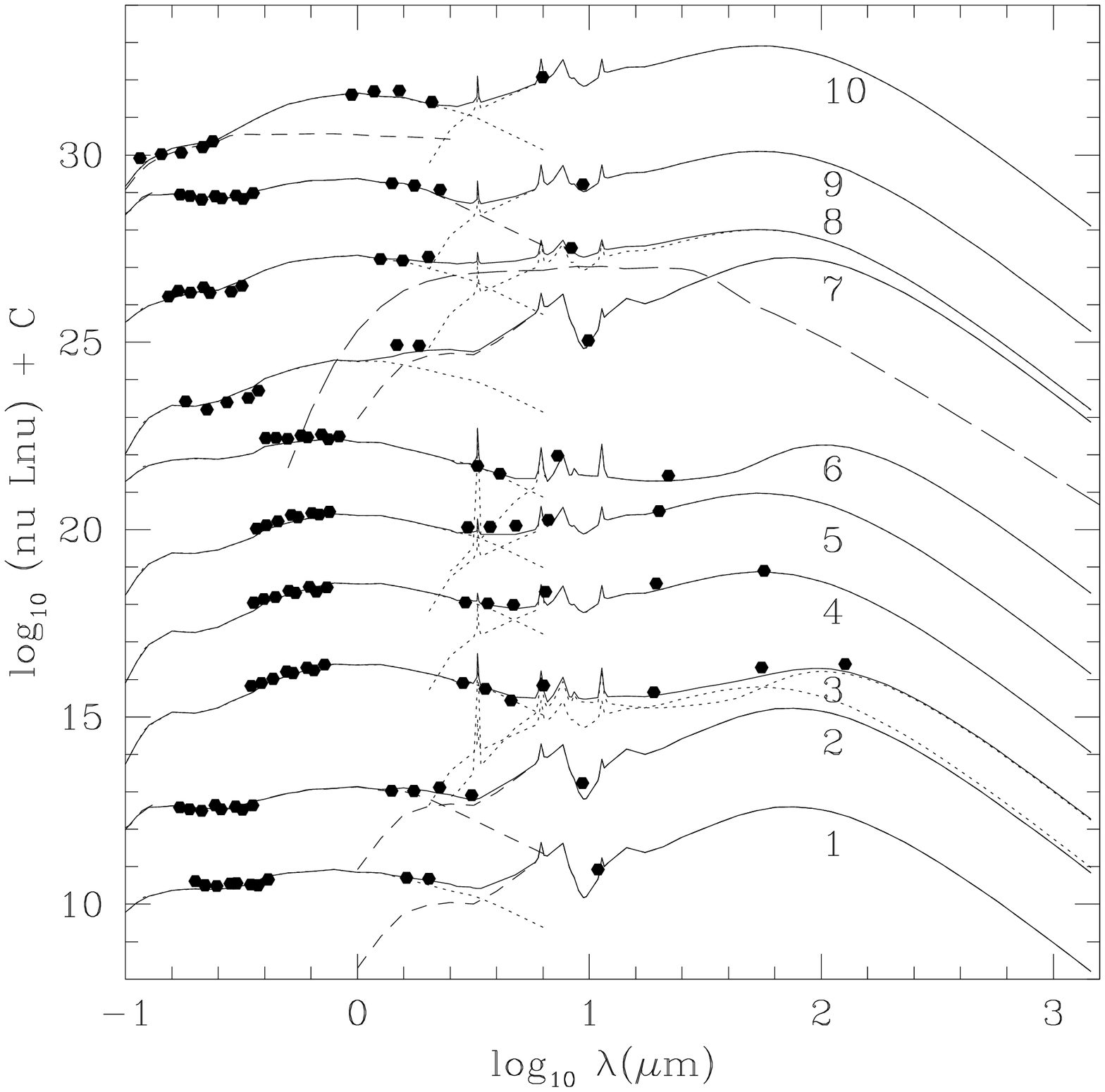,angle=0,width=7cm}
\caption{LH: SEDs of objects with very weak tori (Table 3).
RH: SEDs of candidates for being X-ray starbursts (Table 4).
}
\end{figure*}

\begin{table*}
\caption{X-ray sources with weak dust tori}
\begin{tabular}{llllllllllll}
object no. & RA & dec & z & spec. type & $L_{xhc}$ & $log_{10} N(H)$ & $L_{sb}$ & $L_{tor}$ & $L_{opt}$ & type & $A_V$\\
&&&&&&&&&&&\\
1 & 158.81332 & 57.51605 & (0.959) &  & 12.39 & 23.41 & 11.75 & 10.60 & 11.44 & Sbc & 0.4\\
2 & 158.73761 & 57.55341 & 0.937 & 3 & 10.93 & 19.57 & 11.22 & 10.47 & 11.32 & Scd & 0.2\\
&&&&&&&&& (11.95 & QSO & 1.4)\\
3 & 158.74062 & 57.63224 & (0.950) & & 11.50 & 23.07 & 11.61 & 9.52 & 11.14 & Scd & 1.4\\
4 & 158.66597 & 57.63459 & 0.503 & 4 & 9.74 & 19.57 & 10.56 & 10.04 & 10.74 & Scd & 0.2\\ 
&&&&&&&&& (11.41 & QSO & 1.3)\\
5 & 158.45320 & 57.69691 & (0.282) & & 9.69 & 21.28 & 9.79 & 8.74 & 9.49 & Sbc & 0.6\\ 
&&&&&&&&& (10.45 & QSO & 2.4)\\
6 & 158.66617 & 57.73209 & 0.452 & 1 & 11.37 & 21.40 & 11.28 & 10.43 & 11.13 & Sbc & 0.0\\
&&&&&&&&& (12.00 & QSO & 1.6)\\
7 & 158.99966 & 57.73656 & 0.516 & 2 & 11.94 & 22.81 & & 10.15 & 10.85 & Sbc & 0.3\\
8 & 158.69876 & 57.83265 & 1.015 & 3 & 11.81 & 22.76 & 11.93 & 10.03 & 11.43 & Sab & \\
9 & 158.88411 & 57.94242 & 0.947 & 1 & 11.87 & 23.12 & 11.12 & 10.37 & 10.72 & E & \\
&&&&&&&&& 10.4 & QSO1 & \\
10 & 158.35919 & 58.02242 & 0.736 & 4 & 11.55 & 21.09 & 11.72 & 10.75 & 11.46 & Scd & \\ 
&&&&&&&&& (11.91 & QSO1 & 0.7)\\
\end{tabular}
\end{table*}

\section{The AGN-starburst connection and X-ray starbursts}

X-ray emission from starbursts, due to associated X-ray binaries, supernova remnants and diffuse hot gas,
have been studied by Griffiths and Padovani (1990), David et al (1992), Nandra et al (2002), and
Ranalli et al (2003).  Ranalli et al (2003) use 23 local starburst galaxies to derive a mean ratio of far infrared
to X-ray luminosity for starbursts.  Nandra et al (2002) show that this ratio is very similar for z $\sim$ 3 
Lyman-break galaxies.  

Figure 7R shows $L_{sb}$ versus $L_X$ for our SWIRE-CLASXS sample.  We have included A220 and cirrus dominated
sources in this plot, with $L_{ir}$ replacing $L_{sb}$.  The range of ratios derived by 
Ranalli et al (2003) for X-ray emission from starbursts is shown as a pair of broken lines. Five sources
lie within this range, with a few others just below the lower edge.  Fig 12R shows SEDs for sources 
with $log_{10} L_{ir} > log_{10} L_{xh,c} + 1.3$ which have 24 $\mu$m detection (properties given in Table 4).
Apart from object 8, which shows some evidence for dust torus emission, all the other objects are good
candidates to be X-ray starbursts.  All except object 4 would be classified as unabsorbed X-ray sources.
While 4 of the 10 objects have low X-ray luminosities ($L_{Xh} < 10^{42} erg s^{-1}$), the rest are of higher luminosity.
Three are hyperluminous infrared galaxies and have correspondingly strong X-ray emission. It appears that X-ray starbursts can 
have X-ray luminosities ranging up to $10^{44} erg s^{-1}$. Spectroscopically object 1 is classifed as broad-line and
object 5 is classified as narrow-line, but neither shows evidence for an optical QSO or for a dust torus.  We conclude 
that at least 7 of these objects are X-ray starbursts, with objects 1, 5 and
8 showing some evidence for the presence of an AGN.  It would be valuable, especially for the starburst candidates
with z$>$1, to obtain spectroscopic redshifts for them.

For the rest of the CLASXS sample the X-ray emission is presumably due to an AGN.  The strong correlation in Fig 7R is 
therefore interesting, suggesting
a common gas feeding mechanism for both starbursts and for black hole accretion, presumably due
to galaxy interactions and mergers.  Since $L_{sb}$ is proportional to the star formation rate, and for 
a constant Eddington factor, $L_{Xh}$ is proportional to the black hole accretion rate, this plot can be seen as the
time derivative of the Magorrian relation between bulge stellar mass and black hole mass.
An earlier version of this correlation, in the form
$L_{sb}$ versus $L_{tor}$, was given by Rowan-Robinson (2000).

\begin{table*}
\caption{X-ray starburst candidates}
\begin{tabular}{lllllllllllll}
object no. & RA & dec & z & spec. type & $L_{xhc}$ & $log_{10} N(H)$ & $L_{cirr}$ & $L_{sb}$ & $L_{tor}$ & $L_{opt}$ & type & $A_V$\\
&&&&&&&&&&&&\\
1 & 158.67293 & 57.54453 & 1.208 & 4 & 11.10 & 21.65 & & 12.68 (A220) &  & 11.12 & Scd & 0.0 \\
2 & 158.67242 & 57.89325 & (1.570) & & 11.99 & 19.57 & & 13.34 (A220) & & 11.60 & sb & 0.4\\
3 & 158.49472 & 57.72141 & 0.264 & & 9.71 & 21.61 & 11.16 & 10.81 & & 11.64 & Scd & 0.9\\
4 & 158.85077 & 57.74316 & 0.234 & 1 & 9.68 & 22.27 & & 10.87 & & 10.89 & Scd & 0.9\\
5 & 158.54678 & 57.92445 & 0.203 & 2 & 9.08 & 21.04 & & 10.96 &  & 10.56 & Scd  & 0.6\\
6 & 158.32266 & 57.87665 & 0.094 & 1 & 8.07 & 19.57 & 9.16 & & & 9.40 & Scd & 0.0 \\
7 & 158.14431 & 57.93956 & (1.432) &  & 10.94 & 19.57 & & 13.36 (A220) & & 10.91 & Scd & 0.8\\ 
8 & 158.68903 & 57.67629 & (1.871) & & 10.68 & 19.57 & & 12.23 & 11.33 & 11.49 & Sbc & 0.0\\
9 & 158.21892 &57.74105 & (1.559) & & 10.74 & 19.57 & & 12.30 & & 11.82 & sb & 0.3\\
10 & 158.71025 & 57.68777 & (2.819) & & 11.29 & 19.57 & & 13.17 & & 11.92 & Sbc & 0.4\\
 & & & & & & & & & & (11.5 & QSO & 0.7) \\
\end{tabular}
\end{table*}

\section{Discussion and conclusions}

As has been argued in several recent papers (Polletta et al 2006, Daddi et al 2007, Fiore et al 2008, 2009),
combining data from X-ray surveys and from Spitzer gives us a better understanding of the
statistics of the dust-covering factor around AGN.  We have combined the well-studied CLASXS Chandra survey
in Lockman with the SWIRE survey data in the overlapping area of the sky, 0.4 sq deg.  The sample consists
of 401 X-ray-sources, of which 306 are detected by Spitzer, and a further 257 AGN detected through their dust
torus, but not by Chandra.  We have used the spectroscopic redshifts and classifications of Steffen et al (2004), and
other spectroscopy from the literature, where available, and photometric redshifts from the methodology of
Rowan-Robinson et al (2008) for the remainder.  For X-ray sources the X-ray hardness ratio has been modelled 
in terms of a power-law ($\Gamma$ = 1.9) with absorption N(H).  The optical and infrared data have been modelled 
in terms of the galaxy and QSO templates, and infrared templates, of Rowan-Robinson et al (2008).  
The present analysis in terms of well-established infrared templates based on radiative transfer models gives better 
insight into the infrared SEDs, and a better separation of the contribution of starbursts and AGN dust tori, than a simple 
comparison of 24 $\mu$m to optical or X-ray fluxes.  We also believe this gives more insight than using a library
of fixed UV-infrared templates. 

Our estimate of the N(H) distribution is consistent with other studies, but we do find a higher proportion of low absorption 
objects at z $<$ 0.5 (58$\%$ of the X-ray sample)  than at z $>$ 0.5 (37$\%$).  X-ray selection effects make it difficult 
to get a good estimate of the total number of high 
extinction AGN, especially at z $<$ 2.  Almost half of the AGN dust tori which we detect with Spitzer but not with 
Chandra are likely to be Compton thick objects and we estimate that at least 20$\%$ of our combined AGN sample 
are Compton-thick objects.  If all the sources detected with Spitzer but not Chandra were Compton thick objects, which
we regard as highly unlikley, the overall percentage of Compton-thick objects would be increased to 39$\%$.
We could not however rule out a model in which the Compton-thick fraction went from 20$\%$ at $L_{Xh} = 10^{44.5}$ erg s$^{-1}$ 
to 50$\%$ at $L_{Xh} = 10^{42.5}$ erg s$^{-1}$, and such a model might provide a viable model of the X-ray background
(Gilli et al 2007).  

A further uncertainty for the X-ray sample arises from the possible presence of enhanced soft X-ray emission
(Nandra and Pounds 1994, Maloney and Reynolds 2000, Ross and Fabian 2005,  Gierlinski and Done 2006,
Nandra and Iwasawa 2007, Winter et al 2008), which can mask the presence of Compton-thick objects.  We suggest that
the objects in Table 2b fall into this category.

\subsection{Comparison with other work on Compton thick AGN}

Polletta et al (2006) identify 7 X-ray sources with N(H) $> 10^{24} cm^{-2}$, estimate a total of 55 in 0.6 sq deg of 
Chandra/SWIRE area, only 20$\%$ of which are detected in X-rays.  While we find only one X-ray AGN with
N(H) $> 10^{24} cm^{-2}$, our estimate of the total number of Compton-thick objects in our 0.4 sq deg area, to 
comparable Chandra depth, is $\ge$ 130, corresponding to $\ge 20\%$ of the combined SWIRE-CLASXS sample.

Daddi et al (2007) note that 20-30$\%$ of faint z= 1.4-2.5 galaxies detected with MIPS at 24 $\mu$m have mid-infrared 
excess unlikely to be due to star formation, stacked X-ray spectra rising steeply at $>$ 10 KeV, suggesting they host 
Compton thick AGN, and have space-densities twice that of X-ray detected AGN.  This is very consistent with
our estimate of $\ge 20\%$ for a sample most of whose redshifts range from 0.4-3.
 
Fiore et al (2008) studied a population of objects in Chandra DFS with high 24 $\mu$m to optical ratios and estimated that 80$\%$ likely 
to be Compton thick, with the number at z = 1.2-2.6 similar to the number of unobscured and moderately obscured AGNs.
Fiore et al (2009) estimate that 44$\%$ of X-ray sources  with $log_{10} L_{Xh}$ =44-45, and 67$\%$ with $log_{10} L_{Xh}$ = 43.5-44, are Compton 
thick, from COSMOS MIPS survey.  These are much higher percentages than we find (see Table 1).

Alexander et al (2008) note the rarity of reported Compton-thick objects at z = 2-2.5.  They discuss a z = 2.211 Compton thick 
quasar for which they have optical and IRS spectroscopy and identifiy a further 6 at z = 2-2.5 in the literature.  The present study
identifies a further 4 X-ray selected Compton-thick objects at z $>$ 2 on the basis of photometric redshifts (Table 2), and
there are a further 6 infrared-detected AGN at z $>$ 2 which we estimate to be Compton-thick.  Clearly it is desirable to obtain
optical spectroscopy for these Compton-thick candidates.

\subsection{AGN with weak tori, X-ray starbursts}
We conclude that there is no evidence for AGN with no dust tori, and none with a covering factor $<1\%$.  
The range of dust covering factors is 1-100 $\%$, with a mean of 40$\%$, ie a Type 2 fraction of 40$\%$.
However dust covering factors of just a few percent
are not very different from naked quasars and these objects with very weak tori do not fit easily
into a unification model in which AGN properties depend only on viewing angle.
  
Measured by the ratio of dust torus luminosity to X-ray or (for Type 1 objects) optical luminosity, the covering factor
appears to decrease towards lower AGN luminosity, and certainly shows no evidence of an increase,
in contradiction to estimates based on ratios of narrow-line and broad-line spectra. 
The lines-of-sight to X-ray emitting, line emitting and optical continuum emission are complex and 
often involve scattering processes.  The material responsible for the soft X-ray absorption is not necessarily well
correlated with the dust responsible for the dust torus emission.

We find 7-10 X-ray starbursts in the SWIRE-CLASXS sample, with X-ray luminosities ranging up to $L_{Xh} = 10^{44}$ erg s$^{-1}$.
This is a considerable extension of the luminosity range of X-ray starbursts previously reported, but is consistent 
with the an extrapolation of the X-ray-infrared relation for starbursts into the realm of hyperluminous infrared galaxies.

\subsection{Reliability of different luminosities}
Unless we happen to have 70 $\mu$m detections, estimates of starburst luminosities, and of infrared bolometric luminosities
from {\it Spitzer} data are uncertain by a factor of $\sim$ 2 (Rowan-Robinson et al 2005).  Because we are sampling
the AGN dust torus template at its peak wavelengths, and often have several detected bands, the estimates of 
dust torus luminosity are more accurate, with an uncertainty $\sim \pm$ 0.1 dex.  The starburst and dust torus templates 
are so different at 3-24 $\mu$m that there is little aliasing between them even with our minimum two detected
infrared bands.

What is the best way to estimate the total (X-ray to 3 $\mu$m) bolometric luminosity of AGN ?
For Type 1 objects the best estimate is undoubtedly from the extinction corrected optical luminosity, since this involves only
a short extrapolation to the UV peak of the spectral energy distribution.  We have used
the optical-UV bolometric correction of Rowan-Robinson et al (2008), to give $L_{bh} \sim  2.0 L_{opt}$, with
 an uncertainty of $\pm$0.1 dex.

For Type 2 objects, or for Seyferts where the AGN is outshone in the optical by the parent galaxy, the absorption-corrected
hard X-ray luminosity gives a less reliable estimate.  We find a mean X-ray bolometric correction of 27, but with a big range, 
consistent with the 4-100 range found by Vasudevan and Fabian (2007).  We estimate the rms uncertainty in  the X-ray bolometric 
correction to be 0.4 dex.
  
For both Type 1 and Type 2 AGN we can use the luminosity in the AGN dust torus, assuming an average covering factor of
40$\%$.  The range of covering factors is 0.01-1, and the rms uncertainty in the estimate $L_{bh} \sim 2.5.L_{tor}$ is
0.26 dex (Rowan-Robinson et al 2008).  

Finally it is worth remarking that our modelling of infrared SEDs will be enormously improved when 
{\it Herschel} data become available.

\section{Acknowledgements}
We thank the referee for helpful comments which allowed us to improve the paper.


\end{document}